\documentclass[fleqn,usenatbib]{mnras}

\usepackage{newtxtext,newtxmath}

\usepackage[T1]{fontenc}

\DeclareRobustCommand{\VAN}[3]{#2}
\let\VANthebibliography\thebibliography
\def\thebibliography{\DeclareRobustCommand{\VAN}[3]{##3}\VANthebibliography}

\usepackage{graphicx}	
\usepackage{amsmath}	
\usepackage{tikz}
\usepackage{xcolor}
\usepackage{enumitem}

\definecolor{sbblue}{HTML}{4C72B0}
\definecolor{sborange}{HTML}{DD8452}
\definecolor{sbgreen}{HTML}{55A868}
\definecolor{sbred}{HTML}{C44E52}

\title[Reshaping of radially biased stellar haloes]{GSE vs. LMC: reshaping of radially biased stellar haloes by satellites}

\author[A. M. Dillamore et al.]{
Adam M. Dillamore,\thanks{E-mail: a.dillamore@ucl.ac.uk (AMD)}
Jason L. Sanders
and Richard A. N. Brooks
\\
Department of Physics and Astronomy, University College London, London, WC1E 6BT, UK\\
}

\date{Accepted XXX. Received YYY; in original form ZZZ}

\pubyear{\the\year{}}

\begin{document}
\label{firstpage}
\pagerange{\pageref{firstpage}--\pageref{lastpage}}
\maketitle

\begin{abstract}
Perturbations from the Large Magellanic Cloud (LMC) of the Milky Way's stellar and dark matter haloes are well-established. However, studies have generally not considered the high radial anisotropy of the Milky Way's inner halo caused by the accreted debris of \textit{Gaia} Sausage-Enceladus (GSE). We run a series of test particle simulations of stellar haloes being perturbed by the LMC, with different halo velocity anisotropies $\beta\in[0.5,0.9]$. The LMC causes these initially axisymmetric haloes to become approximately triaxial. Their major axes are aligned with its orbital plane and tilted by up to $\sim14^\circ$ with respect to a fixed Galactic disc. These effects become much more dramatic as $\beta$ increases, causing the halo to fractionate spatially according to anisotropy. This confirms the expectations of an analytical model, which predicts that orbits with eccentricities $e\gtrsim0.95$ should azimuthally align with the tidal field of the LMC. The reshaping of the $\beta=0.9$ halo creates strong overdensities of $\sim40\%$ at heliocentric distances as close as 15\,kpc. These coincide with the well-known Virgo Overdensity (VOD) and Hercules-Aquila Cloud (HAC), which have previously been associated with the GSE. We propose that the HAC and VOD were created by the dynamical alignment of highly eccentric orbits by the LMC, and are not necessarily relics of the GSE merger geometry. We conclude that previous works have significantly underestimated perturbations from the LMC in the inner stellar halo by not considering sufficiently high velocity anisotropy. This effect should be corrected for when constructing equilibrium models of the GSE debris.
\end{abstract}




\section{Introduction}
The \textit{Gaia} mission \citep{gaia} revealed that a large fraction of the Milky Way's stellar halo is made up of a population of stars on highly eccentric orbits \citep{belokurov2018,helmi2018,deason2024}. These were likely deposited in a single merger event $\sim8-11$\,Gyr ago \citep{belokurov2020,bonaca2020,das2020,feuillet2021,laporte2026} with a galaxy of stellar mass $\sim1-7\times10^8\,\mathrm{M}_\odot$ \citep[e.g.][]{helmi2018,naidu2020,feuillet2020,mackereth2020,han2022,lane2023}, known as \textit{Gaia} Sausage-Enceladus (GSE).\footnote{We refer to the ancient galaxy as `GSE', and its present-day remnant in the Milky Way as `GSE debris'.} The kinematics of its stars are typically quantified by the \textit{anisotropy parameter} $\beta$. This is defined by
\begin{equation}\label{eq:beta}
    \beta=1-\frac{\sigma_\theta^2+\sigma_\phi^2}{2\sigma_r^2},
\end{equation}
where $\sigma_i$ is the velocity dispersion in the $i$th direction. $\beta$ therefore equals 0 for a kinematically isotropic distribution, and 1 for a distribution with purely radial orbits. In general $\beta$ is a function of position, particularly radius. Velocity measurements of GSE stars have determined that its anisotropy is as high as $\beta\approx0.9$ within $\sim25$\,kpc of the Galactic centre \citep{lancaster2019,iorio2021,bird2021}.

Various studies have modelled the 3D structure of the GSE debris in the Milky Way. Both \citet{han2022} and \citet{lane2023} found that the distribution is triaxial, with the long axis tilted out of the Milky Way's disc plane. Other works have similarly concluded that the stellar halo as a whole is tilted with respect to the plane \citep{iorio2019,lucey2025,li2025}. Given the proposed merger time, some studies have interpreted this tilt as evidence that the dark matter halo of the Milky Way must also be tilted \citep{han2022b,han2023}. In this scenario it is proposed that the non-axisymmetric structure of the GSE debris has been preserved without being mixed for $\gtrsim8$~Gyr. In \citet{dillamore2025c} we fitted equilibrium dynamical models to the triaxial tilted GSE debris distribution reported by \citet{han2022}, in order to constrain the geometry of the dark matter halo consistent with steady state. We found that a tilted aspherical dark matter halo is required for equilibrium.

The halo also contains smaller-scale structures that are associated with the GSE. The Virgo Overdensity \citep[VOD;][]{vivas2001,newberg2002} and Hercules-Aquila Cloud \citep[HAC;][]{belokurov2007} are overdensities of stars above and below the Galactic plane at distances of $10\lesssim d/\mathrm{kpc}\lesssim25$ \citep{sesar2010,simion2014,grillmair2016}. These are composed of stars on highly eccentric orbits, and are chemodynamically consistent with each other and with the GSE debris \citep{simion2019,balbinot2021,perottoni2022,yan2023,ye2024}. They contribute to the overall tilt of the halo \citep{ye2024}, and it has been proposed that they preserve information about the GSE's merger geometry \citep{naidu2021,han2022b,perottoni2022}. Alternatively, it has been proposed by \citet{donlon2019} that the VOD is the remnant of a more recent merger event, known as the Virgo Radial Merger. This is hypothesised to have occurred less than 3 Gyr ago \citep{donlon2023}, and resulted in debris occupying similar regions to that from the GSE \citep{donlon2019}.

Interpreting the stellar halo is complicated by ongoing perturbations, most notably from the Large Magellanic Cloud (LMC). The LMC is the Milky Way's largest satellite, and has just passed pericentre on its first passage \citep[or possible second; see][]{vasiliev2024} around the Galaxy \citep[see][for a comprehensive review]{vasiliev2023}. Its considerable total mass of $M_\mathrm{LMC}\sim10^{11}M_\odot$ \citep{penarrubia2016,erkal2019,vasiliev_tango,correa_magnus2022,koposov2023} is resulting in significant perturbations to the density and kinematics of the Milky Way's stellar and dark matter haloes. The LMC produces an overdense conical wake (the \textit{transient response}) behind it along its orbit, while a broader overdensity (the \textit{collective response}) is formed on the opposite side of the Galaxy \citep{chandrasekhar1943,weinberg1989,garavito-camargo2019,garavito-camargo2021,conroy2021,rozier2022,foote2023}. The acceleration of the Milky Way's centre towards the LMC causes it to undergo \textit{reflex motion} relative to regions of the halo at large radius. This manifests itself as an apparent bulk motion of halo stars towards the northern Galactic hemisphere (away from the LMC's past orbit), which is observable as a dipole signature in both density \citep[e.g.][]{garavito-camargo2021,conroy2021} and velocities \citep[e.g.][]{peterson2020,erkal2021,yaaqib2024,chandra2025,bystrom2025}. These perturbations have consequences for equilibrium modelling of the halo, and must be corrected for to avoid biased inference of the Milky Way's mass \citep{erkal2020,correa_magnus2022}. Measurements of the halo density and velocity perturbations are extremely useful tools for inferring the mass, structure, and orbit of LMC \citep[e.g.][]{brooks2025b, Brooks2026}, and are complemented by constraints from LMC-induced perturbations of stellar streams \citep{erkal2019,shipp2021,koposov2023}.

Simulations predict that the density changes to the stellar halo depend on its kinematics. \citet{rozier2022} demonstrated that haloes with more radially biased velocity distributions (higher $\beta$) experience stronger density contrasts (particularly around the LMC's wake), and the geometry of these enhancements is a function of $\beta$. However, this study only considered values of $\beta$ up to 0.49, well below the measured value of $\beta=0.9$ for the GSE debris. Similarly \citet{garavito-camargo2019} ran $N$-body simulations of the Milky Way-LMC interaction with values of $\beta$ no higher than $\approx0.6$. From \citet{rozier2022} we may expect the GSE debris to respond much more strongly to the LMC than the simulated haloes in these studies.

In this paper we consider the effects of the LMC on a stellar halo with very high radial anisotropy, comparable to the GSE debris. We investigate how the LMC changes the global shape and orientation of the halo, in order to determine whether it could be responsible for the observed tilted triaxial structure (including the HAC and VOD). We use two models: an analytical toy model describing how highly eccentric orbits behave in the tidal field of the LMC, and a set of test particle simulations. The latter feature realistic stellar haloes with different velocity anisotropies being perturbed by the LMC. Their initial conditions are constructed using the Schwarzschild (orbit-superposition) method \citep{schwarzschild1979}, which we previously used to build equilibrium models of the GSE debris in \citet{dillamore2025c}.

The rest of this paper is arranged as follows. In Section~\ref{section:theory} we present an analytic model describing how satellites such as the LMC perturb highly eccentric orbits. We describe the setup of our simulations in Section~\ref{section:simulations} and present their results in Section~\ref{section:results}. Finally we summarise our conclusions in Section~\ref{section:conclusions}.

\section{Theory}\label{section:theory}
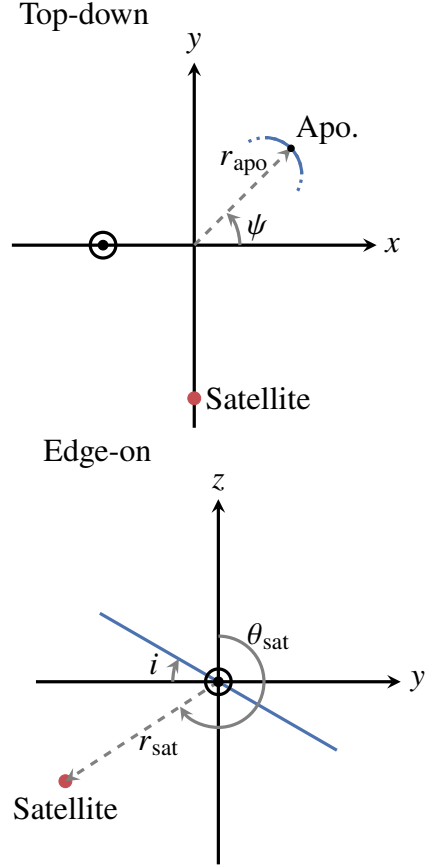
\begin{figure}
    \centering
    \begin{minipage}{0.3\textwidth}
        \centering
        \begin{tikzpicture}[scale=1.2,>=stealth]
            \def\alphahalo{204.33} 
            \def\r{2}
            \def\rapo{1.5}
            \def\phiapo{45}
            \def\thetas{90-237} 
            \def\rs{2}
            
            \draw[->, very thick] (-\r,0) -- (\r,0) node[anchor=west] {\Large $x$};
            \draw[->, very thick] (0,-\r) -- (0,\r) node[anchor=south] {\Large $y$};

            \filldraw[black] (-1,0) circle (1.5pt);
            \draw[black, very thick] (-1,0) circle (4pt);
            
            \node[anchor=west] at (-2,2.5) {\Large Top-down};

            \draw[dotted, very thick, sbblue] ({\rapo*cos(\phiapo)},{\rapo*sin(\phiapo)}) arc (45:120:0.4);
            \draw[dotted, very thick, sbblue] ({\rapo*cos(\phiapo)},{\rapo*sin(\phiapo)}) arc (45:-30:0.4);

            \draw[very thick, sbblue] ({\rapo*cos(\phiapo)},{\rapo*sin(\phiapo)}) arc (45:90:0.4);
            \draw[very thick, sbblue] ({\rapo*cos(\phiapo)},{\rapo*sin(\phiapo)}) arc (45:-0:0.4);

            \draw[->, dashed, very thick, gray] (0,0)--({\rapo*cos(\phiapo)},{\rapo*sin(\phiapo)});
            \node at (0.55,0.9) {\Large $r_\mathrm{apo}$};

            \filldraw ({\rapo*cos(\phiapo)},{\rapo*sin(\phiapo)}) circle (1pt);
            
            \draw[->, very thick, gray] (0.5,0) arc (0:45:0.5);
            \node at (0.7,0.2) {\Large $\psi$};

            \node at ({\rapo*cos(\phiapo)+0.4},{\rapo*sin(\phiapo)+0.2}) {\Large Apo.};

            \filldraw[sbred] (0,{\rs*cos(\thetas)}) circle (2pt);
            \node at (0.7,{\rs*cos(\thetas)}) {\Large Satellite};
            



            
            
                   

            \end{tikzpicture}
            
            \end{minipage}
            \hfill


    \begin{minipage}{0.3\textwidth}
        \centering
        \begin{tikzpicture}[scale=1.2,>=stealth]
            \def\thetas{90-237} 
            \def\r{2}
            \def\rapo{1.5}
            \def\rs{2}
            \def\i{30}
            
            \draw[->, very thick] (-\r,0) -- (\r,0) node[anchor=west] {\Large $y$};
            \draw[->, very thick] (0,-\r) -- (0,\r) node[anchor=south] {\Large $z$};

            \draw[very thick, sbblue] ({-\rapo*cos(\i)},{\rapo*sin(\i)})--({\rapo*cos(\i)},{-\rapo*sin(\i)});

            \draw[->, very thick, gray] (-0.5,0) arc (0:-\i:-0.5);
            \node at (-0.7,0.2) {\Large $i$};
            
            \draw[->, very thick, gray] (0,0.5) arc (90:\thetas:0.5);
            \node at (0.55,0.5) {\Large $\theta_\mathrm{sat}$};

            \filldraw[sbred] ({\rs*cos(\thetas)},{\rs*sin(\thetas)}) circle (2pt);
            \node at ({\rs*cos(\thetas)+0.2},{\rs*sin(\thetas)-0.3}) {\Large Satellite};

            \draw[->, dashed, very thick, gray] (0,0)--({\rs*cos(\thetas)},{\rs*sin(\thetas)});
            \node at (-0.65,-0.7) {\Large $r_\mathrm{sat}$};

            \filldraw[black] (0,0) circle (1.5pt);
            \draw[black, very thick] (0,0) circle (4pt);

            \node[anchor=west] at (-2,2.5) {\Large Edge-on};

        \end{tikzpicture}
    \end{minipage}
    \caption{Geometry of our analytic model setup. \textbf{Top panel:} top-down view of the Galactic plane, with an illustration of an orbit around its apocentre at radius $r_\mathrm{apo}$ and azimuth $\psi$. \textbf{Bottom panel:} edge-on view of the Galactic plane with the location of the satellite marked, at radius $r_\mathrm{sat}$ and polar angle $\theta_\mathrm{sat}$. An example of an orbit with inclination $i$ is also shown.}\label{fig:coords}
\end{figure}

In this section we develop a simple theory to predict the effects of the LMC on a set of highly eccentric orbits in the Milky Way, including the criteria and timescales of the response. We will concentrate here on how the LMC affects the distribution of orbital orientations within the Galactic plane, as opposed to out-of-plane 3D effects.

We represent the Milky Way with a spherical isochrone potential \citep{Henon}, which has the advantage that all orbits are fully analytic \citep{binney_tremaine}. This is defined by
\begin{equation}
    \Phi(r)=-\frac{GM}{b+\sqrt{b^2+r^2}},
\end{equation}
where $M$ and $b$ are the mass and scale length, and $r$ is radius. Following \citet{dillamore2024}, we set $M=2.35\times10^{11}\mathrm{M}_\odot$ and $b=3$\,kpc to approximately match the Milky Way's rotation curve within radii of $\sim20$\,kpc.

We consider orbits perturbed by a tidal force from a satellite in the $y$-$z$ plane at position $\boldsymbol{r}_s=r_\mathrm{sat}\hat{\boldsymbol{r}}_\mathrm{sat}$, where $\hat{\boldsymbol{r}}_\mathrm{sat}\equiv(0,\mathrm{sin}\,\theta_\mathrm{sat},\mathrm{cos}\,\theta_\mathrm{sat})$ (see Fig.~\ref{fig:coords} for the geometry of the setup). For a perturber of mass $m$, the tidal acceleration on a particle at position $\boldsymbol{r}=(x,y,z)$ is
\begin{align}
    \boldsymbol{a}_\mathrm{tide}(\boldsymbol{r})&\approx-\frac{Gm}{r_\mathrm{sat}^3}\left(\boldsymbol{r}-3(\boldsymbol{r}\cdot\hat{\boldsymbol{r}}_\mathrm{sat})\hat{\boldsymbol{r}}_\mathrm{sat}\right)\\
    &=-\frac{Gm}{r_\mathrm{sat}^3}\left(x,(1-3\,\mathrm{sin}^2\,\theta_\mathrm{sat})y,(1-3\,\mathrm{cos}^2\,\theta_\mathrm{sat})z\right),
\end{align}
where we assume that $|\boldsymbol{r}|\ll r_\mathrm{sat}$. These equations result from expanding the difference in the perturber's acceleration between the particle and the origin to leading order in $|\boldsymbol{r}|/ r_\mathrm{sat}$. The tidal force exerts a torque on the particle's orbit, causing the $z$-component of its angular momentum to change at a rate of
\begin{align}
    \dot{L}_z&=\left(\boldsymbol{r}\times\boldsymbol{a}_\mathrm{tide}(\boldsymbol{r})\right)\cdot\hat{\boldsymbol{z}}\\
    &=3\frac{Gm\,\mathrm{sin}^2\,\theta_\mathrm{sat}}{r_\mathrm{sat}^3}xy\\
    &=\frac{3}{2}\Omega_\mathrm{sat}^2R^2\,\mathrm{sin}\,2\phi,\label{Lz_dot}
\end{align}
where $R$ and $\phi$ are the cylindrical Galactocentric distance and azimuth, and we have defined
\begin{align}
    \Omega_\mathrm{sat}^2(r_\mathrm{sat},\theta_\mathrm{sat})&\equiv\frac{Gm\,\mathrm{sin}^2\,\theta_\mathrm{sat}}{r_\mathrm{sat}^3}.
\end{align}
$\Omega_\mathrm{sat}$ has units of frequency and encodes all the parameters associated with the satellite. In general $\Omega_\mathrm{sat}$ is a function of the satellite's position $(r_\mathrm{sat},\theta_\mathrm{sat})$, so varies with time.

We wish to consider how this torque affects the orbit's orientation. For highly eccentric orbits this is best described by the apocentric angle (measured in the orbital plane). Near-radial orbits pass through the centres of galactic potentials on almost straight lines, so that pairs of successive apocentres are separated in angle by $\approx\pi$ radians. We therefore define $\psi$ to be the angle of \emph{alternate} apocentres, so that a non-precessing orbit will have constant $\psi$. This angle evolves according to
\begin{equation}
    \dot{\psi}=\Omega_\theta-\frac{1}{2}\Omega_r\,
\end{equation}
where $\Omega_r$ and $\Omega_\theta$ are the radial and angular frequencies of the orbit \citep{binney_tremaine}. In the isochrone potential these frequencies are related by \citep{binney_tremaine}
\begin{align}
    \Omega_\theta&=\frac{1}{2}\left(1+\frac{L}{\sqrt{L^2+4GMb}}\right)\Omega_r\\
    &=\frac{1}{2}\left(1+\frac{L}{2\sqrt{GMb}}+O\left(\frac{L^2}{4GMb}\right)\right)\Omega_r,
\end{align}
where $L$ is the magnitude of the particle's angular momentum. We consider orbits with small $L/\sqrt{GMb}\sim L/(1740\,\mathrm{kpc\,km\,s^{-1}})$, roughly equivalent to those with small angular momentum compared to the Sun. In this case the apocentric azimuth evolves according to
\begin{align}\label{eq:psi_dot}
    \dot{\psi}&\approx\frac{L_z}{4\sqrt{GMb}\,\mathrm{cos}\,i}\,\Omega_r,
\end{align}
where we have used the orbital inclination $i$ (defined by $\mathrm{cos}\,i\equiv L_z/L$) to replace $L$ with $L_z$. This can be rewritten by noting that the energy in the isochrone potential is $E=-\frac{1}{2}(GM\Omega_r)^{2/3}$ \citep{binney_tremaine}, and by defining a characteristic orbital radius,
\begin{equation}\label{eq:semi-major_axis}
    a\equiv-\frac{GM}{2E}.
\end{equation}
We note that for purely radial orbits, $a$ is related to the apocentric radius $r_\mathrm{apo}$ by
\begin{align}
    a(L=0)&=\frac{1}{2}\left(b+\sqrt{b^2+r_\mathrm{apo}^2}\right)\\
    &\approx\frac{r_\mathrm{apo}}{2}\quad\text{if $r_\mathrm{apo}\gg b$}.
\end{align}
Using the definition of $a$, Equation~\eqref{eq:psi_dot} becomes
\begin{align}\label{eq:psi_dot_simplified}
    \dot{\psi}&\approx\left[{4b^{1/2}a^{3/2}\mathrm{cos}\,i}\right]^{-1}L_z.
\end{align}
We now ignore changes to the inclination by setting $i=\mathrm{constant}$, and only consider the orbit's azimuthal evolution. We differentiate Equation~\eqref{eq:psi_dot_simplified}, combine with Equation~\eqref{Lz_dot}, and average over a radial period to find that the acceleration of the apocentric angle is
\begin{align}
    \ddot{\psi}&\approx\frac{3}{8}\Omega_\mathrm{sat}^2\left[{b^{1/2}a^{3/2}\mathrm{cos}\,i}\right]^{-1}\,\langle R^2\,\mathrm{sin}\,2\phi\rangle,
\end{align}
where $\langle\cdot\rangle$ denotes the radial period average. We now simplify this equation by considering only orbits in the plane $z=0$, for which $i=0$ and $\psi=\phi$ at apocentre. We therefore only consider the evolution in the Galactic plane, not the vertical motion. Due to the weighting by $R^2$ and the large proportion of time spent near apocentre, we can approximate the averaged quantity as,
\begin{equation}
    \langle R^2\,\mathrm{sin}\,2\phi\rangle\approx \langle R^2\rangle\,\mathrm{sin}\,2\psi.
\end{equation}
This gives the approximate equation of motion for $\psi$,
\begin{align}
    \ddot{\psi}&\approx\frac{3}{8}\Omega_\mathrm{sat}^2\frac{\langle R^2\rangle}{b^{1/2}a^{3/2}\,}\,\mathrm{sin}\,2\psi.
\end{align}
This is instantaneously the equation of motion of a pendulum with stable equilibria at $\psi=\pm\pi/2$. Orbits are therefore expected to preferentially align with their apocentres in the same direction as the perturber (i.e. along the $y$-axis). Defining $\theta\equiv2\psi-\pi$, the equation can be rewritten as
\begin{align}
    \ddot{\theta}&+\omega_0^2\,\mathrm{sin}\,\theta=0,\label{eq:pendulum}\\
    \omega_0^2&\equiv\frac{3}{4}\Omega_\mathrm{sat}^2\frac{\langle R^2\rangle}{b^{1/2}a^{3/2}}.\label{eq:frequency}
\end{align}
The quantity $\langle R^2 \rangle$ can be shown to be (see Appendix~\ref{section:mean_R_squared})
\begin{equation}
    \langle R^2 \rangle=\frac{1}{2}\left[(5a-b)(a-b)-\frac{3aL_z^2}{GM}\right],
\end{equation}
so in terms of the the orbital parameters $L_z$ and $a$, the instantaneous frequency is given by
\begin{equation}
    \omega_0^2=\frac{3\Omega_\mathrm{sat}^2(r_\mathrm{sat},\theta_\mathrm{sat})}{8b^{1/2}a^{3/2}}\left[(5a-b)(a-b)-\frac{3aL_z^2}{GM}\right].
\end{equation}
This provides an estimate of the timescale on which orbital orientations respond to the LMC. 

To compute a numerical estimate for $\omega_0$ we need to specify the LMC's parameters via $\Omega_\mathrm{sat}$. We first set $\Omega_\mathrm{sat}$ to the current approximate value for the LMC, by choosing $m=1.5\times10^{11}\mathrm{M}_\odot$, $r_\mathrm{sat}=50$\,kpc, and $\theta_\mathrm{sat}=237^\circ$ \citep{vasiliev_tango,pietrzynski2019,vandermarel2002}. These give $\Omega_\mathrm{sat}\approx1.96$~km\,s$^{-1}$\,kpc$^{-1}$. For a near-radial orbit with $L_z\ll\sqrt{GMa}$ and characteristic radius $a=10$\,kpc (apocentric radius $r_\mathrm{apo}\approx16.7$\,kpc), this corresponds to a timescale of $\tau_\mathrm{align}\sim\omega_0^{-1}\approx0.3$\,Gyr.

A more conservative estimate for $\Omega_\mathrm{sat}$ can be calculated by setting $r_\mathrm{sat}=100$\,kpc and $\mathrm{sin^2\,\theta_\mathrm{sat}}=1/2$, which roughly correspond to their average values over the last 1\,Gyr \citep[see Section \ref{section:lmc} and][]{vasiliev2023}. In this case $\Omega_\mathrm{sat}\approx0.568$~km\,s$^{-1}$\,kpc$^{-1}$, which gives an alignment timescale for the above orbit of $\tau_\mathrm{align}\sim1$\,Gyr. This is similar to the timescale over which this value of $\Omega_\mathrm{sat}$ is valid, so orbits have enough time to respond to the tidal field. We therefore expect that the LMC is capable of causing significant re-alignment of eccentric inner halo orbits with its orbital plane.

We can also estimate a criterion that orbits must satisfy to be affected by the LMC. Clustering of orbits in the minima of the pendulum potential will only occur if the angle $\theta$ is librating rather than circulating. This is because a circulating $\theta$ will spend less time around the minima and more time around the maxima, whereas a librating $\theta$ will oscillate about a minimum. Since the alignment timescale $\omega_0$ is comparable to the LMC interaction timescale, these effects have sufficient time to change the $\theta$ distribution.

An integral of motion of Equation \eqref{eq:pendulum}, assuming constant $\omega_0$, is the pendulum energy, 
\begin{align}
    E_\mathrm{p}\equiv\frac{1}{2}\dot{\theta}^2-\omega_0^2\,\mathrm{cos}\,\theta,
\end{align}
which can be written in terms of the dimensionless quantity,
\begin{align}
    k^2&\equiv\frac{1}{2}\left(1+\frac{E_\mathrm{p}}{\omega_0^2}\right).
\end{align}
The condition for libration of the pendulum is then simply $k^2<1$. We use Equation \eqref{eq:psi_dot_simplified} to replace $\dot{\theta}$ with $L_z$ and obtain
\begin{equation}
    k^2=\frac{1}{2}\left(1-\mathrm{cos}\,\theta+\frac{L_z^2}{3\Omega_\mathrm{sat}^2\,b^{1/2}a^{3/2}}\left[(5a-b)(a-b)-\frac{3aL_z^2}{GM}\right]^{-1}\right).
\end{equation}
This depends on the orbit's angular momentum $L_z$, energy $E$ (via $a$) and orientation $\psi$ (via $\theta$). The minimum value of $k^2$ at a given $(L_z, E)$ is achieved when $\theta=0$. A necessary condition for an orbit's orientation to be aligned by the satellite's tidal field is therefore
\begin{equation}
    k_\mathrm{min}^2\equiv\frac{L_z^2}{6\Omega_\mathrm{sat}^2\,b^{1/2}a^{3/2}}\left[(5a-b)(a-b)-\frac{3aL_z^2}{GM}\right]^{-1}<1.
\end{equation}
\begin{figure}
  \centering
  \includegraphics[width=\columnwidth]{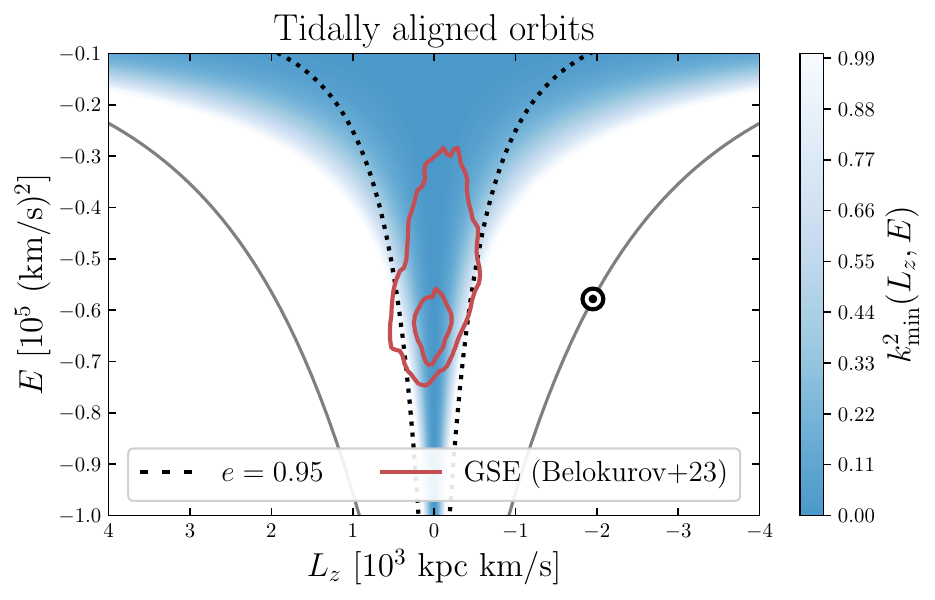}
  \caption{Integral of motion space $(L_z,E)$ coloured by the minimum dimensionless pendulum energy $k_\mathrm{min}^2\equiv k^2(\theta=0)$. The black dotted line indicates orbits with eccentricity $e=0.95$, and the contours show the GSE debris. A circular orbit at the Sun's radius is shown with the $\odot$ symbol. At low energies alignment by the satellite's tidal field is only possible ($k_\mathrm{min}^2<1$) on highly eccentric orbits ($e\gtrsim0.95$).}
   \label{fig:k_squared}
\end{figure}
In Fig.~\ref{fig:k_squared} we show the value of $k_\mathrm{min}^2$ as a function of $L_z$ and $E$. We again set $\Omega_\mathrm{sat}\approx1.96$~km\,s$^{-1}$\,kpc$^{-1}$, its current value for the LMC. The colour increases in darkness as $k_\mathrm{min}^2$ decreases below unity, and white indicates $k_\mathrm{min}^2>1$ (where trapping is not possible). For comparison the dotted black lines indicate orbits with eccentricities of $e=0.95$. Here $e$ is defined by \citep{lynden-bell1963,binney_tremaine}
\begin{align}
    1-e^2&=\left[1+\left|\frac{J_r}{L_z}\right|\right]^{-2} ,
\end{align}
where the radial action
\begin{align}
    J_r&=\sqrt{GMa}-\frac{1}{2}\left(L+\sqrt{L^2+4GMb}\right).
\end{align}
We also show contours of the GSE debris derived by \citet{belokurov_chevrons} from \textit{Gaia} DR3 data, and a circular orbit at the Sun's radius with the $\odot$ symbol.

This shows that at low energies ($E\lesssim E_\odot$), only orbits with high eccentricity ($e>0.95$) are expected to be aligned by the LMC's tidal field. However, this does include a large proportion of stars belonging to the GSE. From Fig.~\ref{fig:k_squared} we therefore expect that the orientations of GSE orbits are significantly affected by the LMC, even at low energy and small radius. At higher energies, the range of $L_z$ in which orbits are affected increases. This is due to a combination of increased torque (via the increase in $\langle R^2\rangle$) and decreased natural precession frequency (via the decrease in $\Omega_r$).

It is worth considering whether the Sagittarius Dwarf Spheroidal Galaxy (Sgr) could be responsible for a similar effect. This is also on a near-polar orbit, roughly orthogonal to that of the LMC \citep[e.g.][]{law2010,vasiliev_tango}. Its remnant has a mass of $M_\mathrm{sat}\approx4\times10^8\mathrm{M}_\odot$ \citep{vasiliev2020_sgr}, and is currently located at $r_\mathrm{sat}\approx19$\,kpc, $\theta_\mathrm{sat}\approx250^\circ$. This gives a corresponding value of $\Omega_\mathrm{sat}\approx0.46$~km\,s$^{-1}$\,kpc$^{-1}$, roughly a quarter of the current value for the LMC. A more conservative estimate with $r_\mathrm{sat}=50$\,kpc \citep[roughly half-way between pericentre and apocentre;][]{vasiliev_tango} and  $\mathrm{sin^2\,\theta_\mathrm{sat}}=1/2$ gives $\Omega_\mathrm{sat}\approx0.08$~km\,s$^{-1}$\,kpc$^{-1}$. Sgr is therefore expected to have a weaker influence on orbital alignments than the LMC. However, we note that Sgr has been orbiting the Milky Way for longer, and its former mass may have been considerably higher than that of the present remnant \citep[e.g.][]{gibbons2017,read2019}. Its effect on the halo may therefore have been greater in the past. We leave a full investigation of the effect of Sgr to a future study.

In this analytic model we have only considered how the orbits align within the Galactic plane $(x,y)$. Since the LMC has passed below the plane of the Milky Way, we may also expect orbital inclinations to be affected. To investigate these 3D effects and test our predictions, we now proceed to run realistic simulations of the stellar halo subject to perturbations from the LMC.
\begin{figure}
  \centering
  \includegraphics[width=\columnwidth]{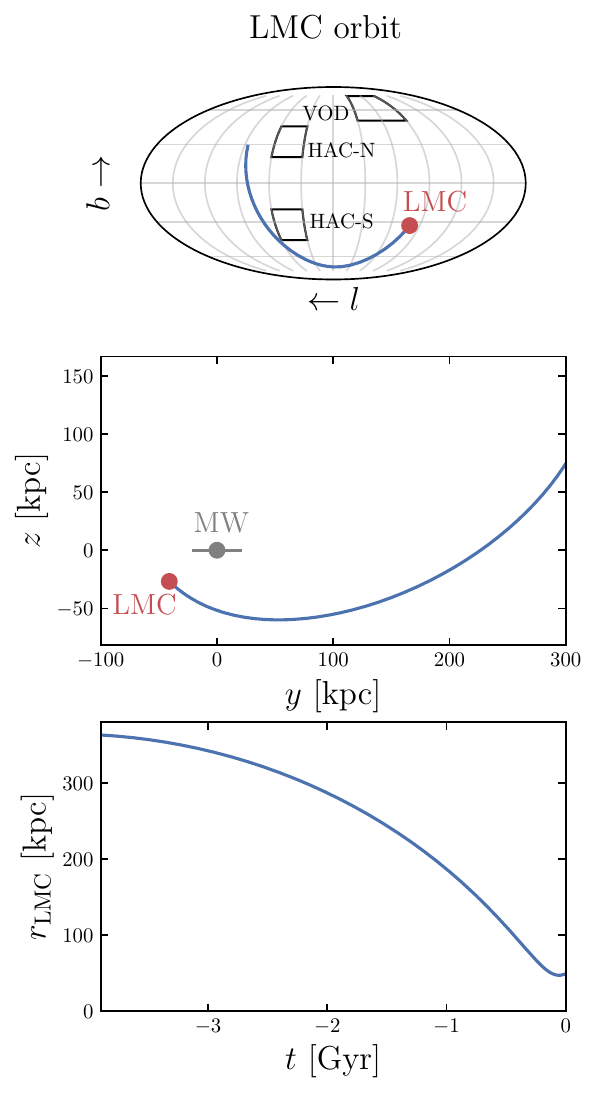}
  \caption{Past orbit of the LMC in our simulations. \textbf{Top panel:} the orbit in Galactic coordinates $(l,b)$ (where $l$ increases to the left). The approximate regions of the HAC and VOD are shown for comparison. The HAC is split into two regions, north (HAC-N) and south (HAC-S) of the Galactic disc (see Section \ref{section:overdensities}). The \textbf{Middle panel:} the track of the orbit in the $(y,z)$ plane, close to the LMC's orbital plane. The current locations of the Milky Way and LMC are marked. \textbf{Bottom panel:} Galactocentric radius $r_\mathrm{LMC}$ of the LMC as a function of time $t$. The simulation runs from $t\approx-4$\,Gyr (when the LMC is near apocentre) to the present-day at $t=0$.}
   \label{fig:lmc_orbit}
\end{figure}
\section{Simulations}\label{section:simulations}
In this section we describe the setup of our simulations of the interaction between the Milky Way's stellar halo and the LMC. The LMC is represented by an analytic rigid potential with no particles, while the stellar halo is modelled as a set of test particles whose self-gravity is ignored. We use the \textsc{agama} software \citep{agama} throughout this section.
\subsection{Milky Way}
The past orbit of the LMC is very sensitive to the choice of potential for the Milky Way \citep{vasiliev2023}. To ensure that this past orbit is realistic, we use the fiducial Milky Way potential from \citet{vasiliev2023}, which consists of a bulge, disc and halo. These respectively have the density profiles,
\begin{align}
    \rho_\mathrm{bulge}&\propto\left(1+r/0.2\,\mathrm{kpc}\right)^{-1.8}\,\mathrm{exp}\left[-(r/1.8\,\mathrm{kpc})^2\right],\\
    \rho_\mathrm{disc}&\propto\mathrm{exp}\left[-R/3\,\mathrm{kpc}-|z|/0.3\,\mathrm{kpc}\right],\\
    \rho_\mathrm{halo}&\propto r^{-1}\left(1+r/14\,\mathrm{kpc}\right)^{-2}\,\mathrm{exp}\left[-(r/300\,\mathrm{kpc})^4\right],
\end{align}
with masses $M_\mathrm{bulge}=1.2\times10^{10}\,\mathrm{M}_\odot$, $M_\mathrm{disc}=5\times10^{10}\,\mathrm{M}_\odot$, and $M_\mathrm{halo}=0.95\times10^{12}\,\mathrm{M}_\odot$. This potential has a circular speed of $233\,$km\,s$^{-1}$ at a radius of $8$\,kpc \citep{mcmillan17}, and an enclosed mass within 50\,kpc of $M(r<50\,\mathrm{kpc})=4.1\times10^{11}\,\mathrm{M}_\odot$. We use the standard right-handed Galactocentric coordinate system in which the Sun has position $\boldsymbol{x}_\odot=(-8.122,0,0.0208)\,$kpc and velocity $\boldsymbol{v}_\odot=(12.9, 245.6, 7.78)\,$km\,s$^{-1}$ \citep{schonrich2010,astropy:2013,astropy:2018,gravity2018,bennett2019}.
\subsection{LMC}\label{section:lmc}
Following \citet{vasiliev_tango}, the LMC's potential $\Phi_\mathrm{LMC}$ is sourced by a truncated Navarro-Frenk-White density profile \citep{NFW},
\begin{equation}
    \rho_\mathrm{LMC}(r)\propto r^{-1}\left(1+r/10.8\,\mathrm{kpc}\right)^{-2}\,\mathrm{exp}\left[-(r/108\,\mathrm{kpc})^2\right].
\end{equation}
We set the total mass to $M_\mathrm{LMC}=1.5\times10^{11}\,\mathrm{M}_\odot$, consistent with \citet{erkal2019}. This model satisfies observational constraints on the LMC's enclosed mass profile by \citet{vandermarel2014}, as shown by fig. 3 in \citet{vasiliev_tango}.

We set the right ascension and declination of the LMC to $\alpha_\mathrm{LMC}=81.28^\circ$ and $\delta_\mathrm{LMC}=-69.78^\circ$ \citep{vandermarel2002}, with associated proper motions $\mu_{\alpha,\mathrm{LMC}}=1.858\,\mathrm{mas\,yr}^{-1}$ and $\mu_{\delta,\mathrm{LMC}}=0.385\,\mathrm{mas\,yr}^{-1}$ \citep{luri2021}. We take the distance and line-of-sight velocity to be 49.6\,kpc \citep{pietrzynski2019} and 262.2\,km\,s$^{-1}$ \citep{vandermarel2002}, respectively.

We transform these coordinates to the Galactocentric frame and integrate the LMC's orbit back in time. We consider the LMC and Milky Way as a two-body system, subject to an additional dynamical friction force acting only on the LMC. The positions and velocities of the Milky Way and LMC evolve according to the coupled equations \citep{vasiliev_tango},
\begin{align}
    \dot{\boldsymbol{x}}_\mathrm{MW}&=\boldsymbol{v}_\mathrm{MW},\label{eq:x_dot_MW}\\
    \dot{\boldsymbol{v}}_\mathrm{MW}&=-\nabla\Phi_\mathrm{LMC}(\boldsymbol{x}_\mathrm{MW}-\boldsymbol{x}_\mathrm{LMC}),\\
    \dot{\boldsymbol{x}}_\mathrm{LMC}&=\boldsymbol{v}_\mathrm{LMC},\\
    \dot{\boldsymbol{v}}_\mathrm{LMC}&=-\nabla\Phi_\mathrm{MW}(\boldsymbol{x}_\mathrm{LMC}-\boldsymbol{x}_\mathrm{MW})+\boldsymbol{a}_\mathrm{DF}.\label{eq:v_dot_LMC}
\end{align}
The dynamical friction acceleration $\boldsymbol{a}_\mathrm{DF}$ is given by the Chandrasekhar formula \citep{chandrasekhar1943},
\begin{align}
    \boldsymbol{a}_\mathrm{DF}&=-\frac{4\pi G^2 M_\mathrm{LMC}\,\rho_\mathrm{MW}(\boldsymbol{x}_\mathrm{LMC})\,\mathrm{ln}\,\Lambda}{|\boldsymbol{v'}_\mathrm{LMC}|^3}\left[\mathrm{erf}(X)-\frac{2X}{\sqrt{\pi}}\mathrm{e}^{-X^2}\right]\boldsymbol{v}'_\mathrm{LMC},\\
    X&\equiv\frac{|\boldsymbol{v}'_\mathrm{LMC}|}{\sqrt{2}\sigma_\mathrm{MW}},\\
    \boldsymbol{v}'_\mathrm{LMC}&\equiv\boldsymbol{v}_\mathrm{LMC}-\boldsymbol{v}_\mathrm{MW},
\end{align}
where $\rho_\mathrm{MW}$ and $\sigma_\mathrm{MW}$ are the density and velocity dispersion of the Milky Way at the LMC's location, and $\mathrm{ln}\,\Lambda$ is the Coulomb logarithm. Following \citet{correa_magnus2022} this is set to $\mathrm{ln}\,\Lambda=\left[\mathrm{ln}(|\boldsymbol{x}_\mathrm{LMC}-\boldsymbol{x}_\mathrm{MW}|/2r_\mathrm{scale})\right]^{1/2}$. The Milky Way's velocity dispersion $\sigma_\mathrm{MW}$ is estimated from a \textsc{quasispherical} distribution function \citep{agama} fitted to its halo component in a spherically averaged version of the potential $\Phi_\mathrm{MW}$. This prescription for the dynamical friction is demonstrated by \citet{correa_magnus2022} to match $N$-body simulations reasonably well (see their section 3.2).

We integrate Equations~\eqref{eq:x_dot_MW}-\eqref{eq:v_dot_LMC} back in time from the present day $t=0$ to $t_0\approx-3.9\,$Gyr. At this time the LMC is more than 350\,kpc from the Milky Way's centre, so its perturbations on the inner stellar halo are neglibible. We transform to a non-inertial reference frame centred on the Milky Way, incorporating the Milky Way's reflex motion towards the LMC as an extra term in the potential $\Phi_\mathrm{reflex}$ giving spatially uniform acceleration. The total potential is then the sum $\Phi=\Phi_\mathrm{MW}+\Phi_\mathrm{LMC}+\Phi_\mathrm{reflex}$. The LMC's orbit relative to the Milky Way is illustrated in Fig.~\ref{fig:lmc_orbit}. The orbit is shown in on-sky Galactic coordinates in the top panel. For comparison we show the locations of the Hercules-Aquila Cloud (HAC) and Virgo Overdensity (VOD); see Table~\ref{tab:overdensities} for their definitions. The middle panel shows the orbital track in the $y$-$z$ plane, and the bottom panel shows the distance from the Milky Way as a function of time.
\begin{figure*}
  \centering
  \includegraphics[width=\textwidth]{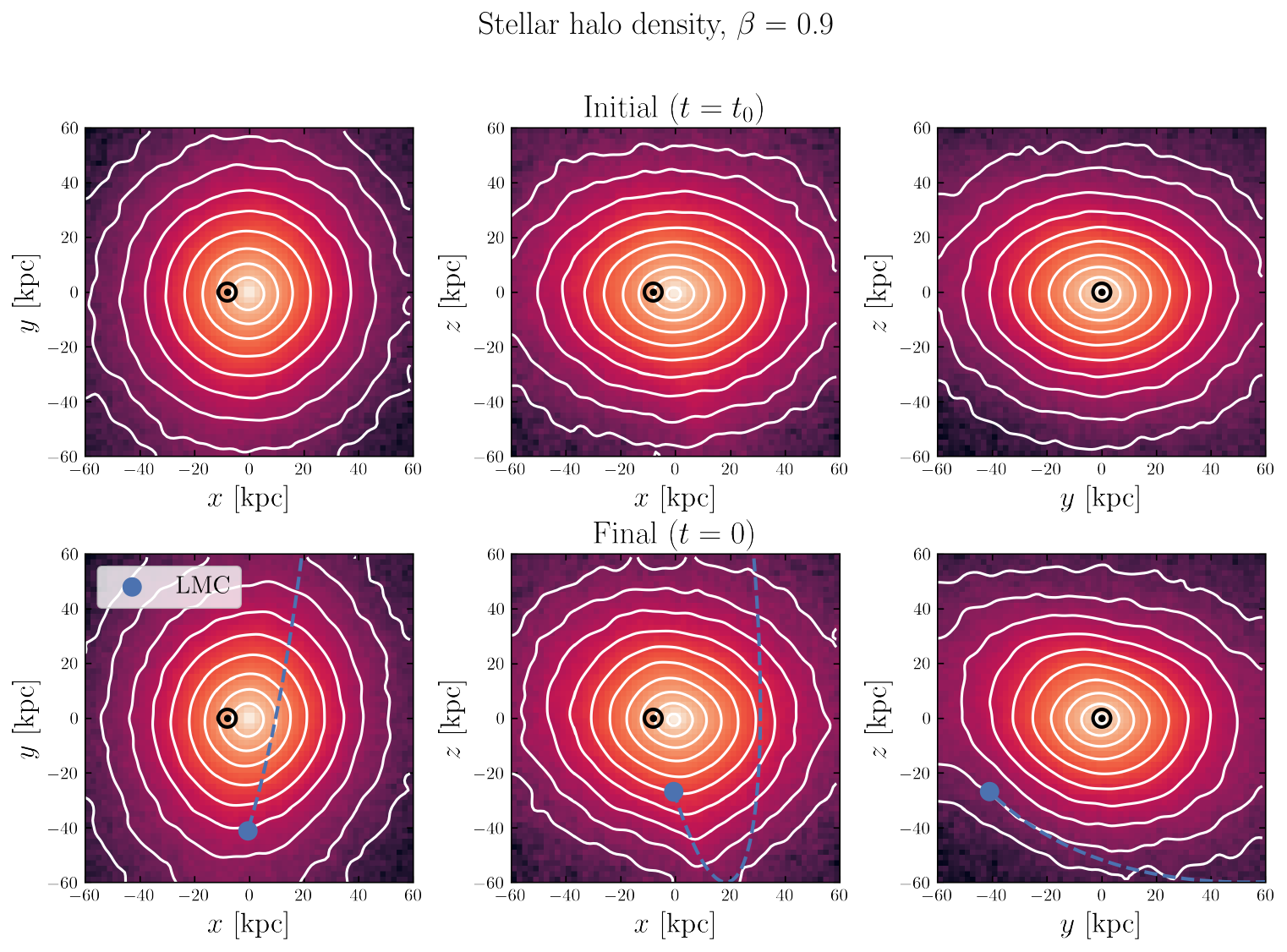}
  \caption{Projected density of our fiducial simulation (with $\beta=0.9$), in the initial (top row) and final (bottom row) snapshots. The left-hand panels show the top-down view of the Galactic plane, and the others show edge-on projections. The LMC and its past orbit are marked, and the location of the Sun is shown with a $\odot$ symbol. The LMC causes the halo to become triaxial and tilted with respect to the Galactic plane.}
   \label{fig:density_3d}
\end{figure*}
\begin{figure*}
  \centering
  \includegraphics[width=\textwidth]{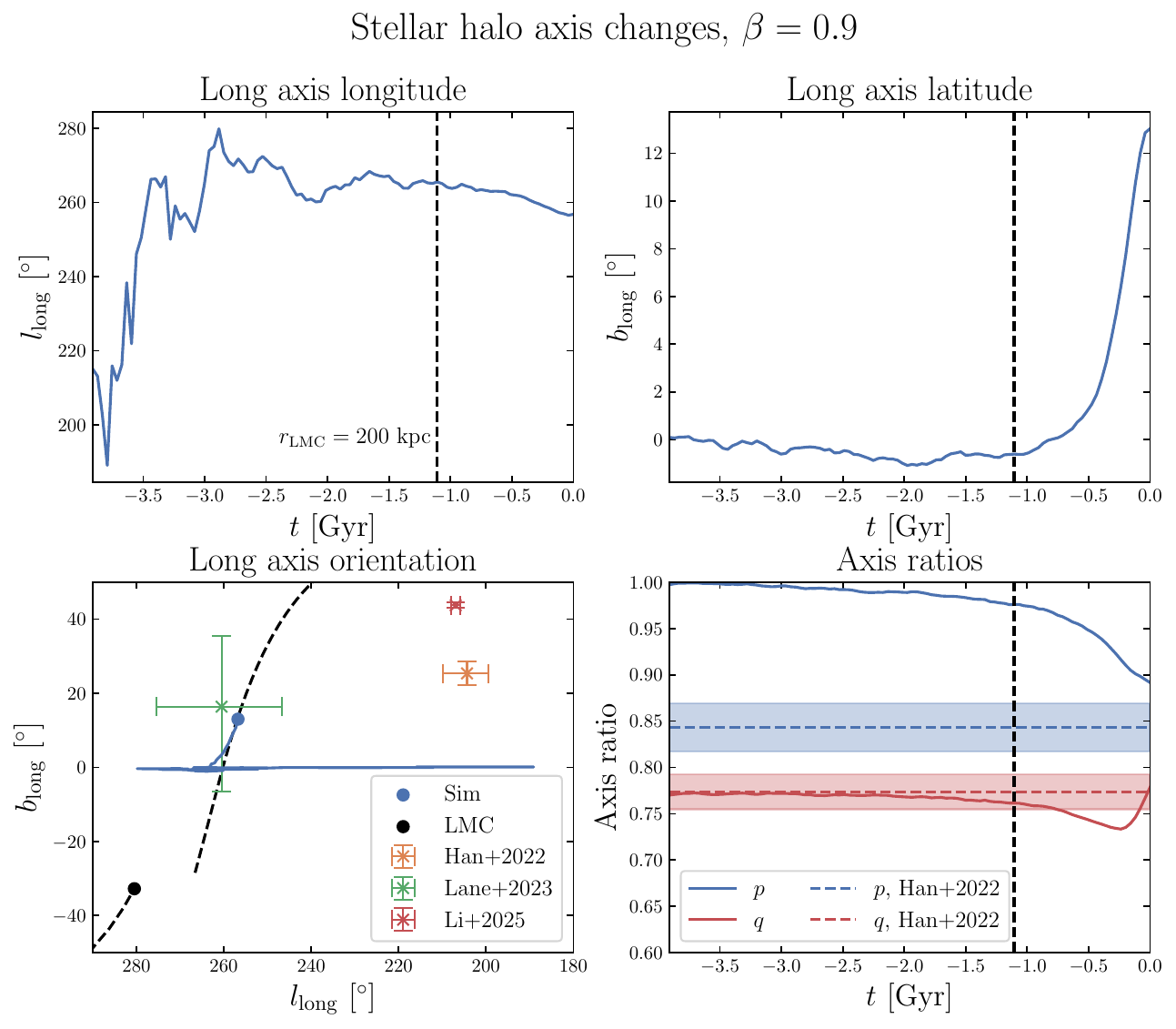}
  \caption{Evolution of the axis ratios and orientations of the $\beta=0.9$ stellar halo. \textbf{Top row:} the Galactic longitude $l_\mathrm{long}$ and latitude $b_\mathrm{long}$ of the halo's long axis as a function of time, where we have taken $180^\circ\leq l_\mathrm{long}<360^\circ$. The vertical dashed lines mark the time at which the LMC is 200 kpc from the Milky Way. \textbf{Bottom-left panel:} the track of the long axis in Galactic coordinates $(l_\mathrm{long},b_\mathrm{long})$ compared to the direction of the LMC and several measurements of the stellar halo. These come from \citet{han2022}, \citet{lane2023}, and \citet{li2025}. The orientation of our simulated halo is very close to the measurement of the GSE debris by \citet{lane2023}. \textbf{Bottom-right panel:} the intermediate-to-long and short-to-long axis ratios $p$ and $q$ as a function of time. The corresponding parameters for the \citet{han2022} fit are shown with dashed lines, with $1\sigma$ uncertainties indicated by coloured bands.}
   \label{fig:axis_changes}
\end{figure*}
\subsection{Stellar halo}
Various methods are available for constructing equilibrium distributions of stars, typically involving action-based distribution function models \citep[e.g.][]{agama}. However, these do not easily allow the kinematics to be changed while keeping the density distribution exactly fixed. We therefore construct stellar halo distributions by Schwarzschild modelling \citep{schwarzschild1979}, which allows us to simultaneously satisfy constraints on the density distribution and velocity anisotropy profile. We use a method similar to that described in \citet{dillamore2025c}. The density distribution of the halo is axisymmetric and has a doubly-broken power-law profile,
\begin{align}\label{eq:rho_dpl}
    \rho_*(r_*)&\propto\begin{cases}
        (r_*/r_\mathrm{b,1})^{-\gamma_1} & r_* \leq r_\mathrm{b,1} \\
        (r_*/r_\mathrm{b,1})^{-\gamma_2} & r_\mathrm{b,1} < r_* \leq r_\mathrm{b,2} \\
        (r_\mathrm{b,2}/r_\mathrm{b,1})^{-\gamma_2}\,(r_*/r_\mathrm{b,2})^{-\gamma_3} & r_* > r_\mathrm{b,2},
    \end{cases}\\
    r_*&\equiv\sqrt{x^2 + y^2 + (z/q_*)^2}.\label{eq:r_*}
\end{align}
We take the break radii and power-law slopes from the fit to the GSE debris by \citet{han2022}. These are respectively $(r_{\mathrm{b},1},r_{\mathrm{b},1})=(11.85,28.33)$\,kpc and $(\gamma_1,\gamma_2,\gamma_3)=(1.70,3.09,4.58)$. We also set the flattening $q_*$ to be equal to the short-to-long axis ratio of the \citet{han2022} fit, $q_*=0.73$. Our stellar halo density distribution is therefore similar to the GSE debris, but initially axisymmetric and aligned with the Galactic plane. However, perturbations from the LMC will cause this shape to change.

We also control the kinematics of the stellar halo by fixing its velocity anisotropy $\beta$ (see Equation~\ref{eq:beta}). Following \citet{dillamore2025c} we set $\beta$ to a constant value in the radial range $6\leq r<24$\,kpc. At larger radii the GSE debris's anisotropy is much less well-constrained \citep[e.g.][]{iorio2021}, so we allow $\beta$ to be free there. For our fiducial model we choose $\beta=0.9$ across this range, to match constraints on the GSE debris \citep{lancaster2019,iorio2021,bird2021,dillamore2025c}. However, we also generate models with a range of $\beta$ between 0.5 and 0.9. This allows us to investigate the anisotropy dependence of our simulations.

We generate the \textit{orbit library} of the Schwarzschild model by sampling $10^4$ particles from the density distribution in Equation~\eqref{eq:rho_dpl}. The velocities are sampled from a Gaussian distribution with dispersion determined by a spherical Jeans model with anisotropy $\beta$. We integrate the orbits of these particles for time $100T_\mathrm{circ}$ in the Milky Way's potential $\Phi_\mathrm{MW}$, where $T_\mathrm{circ}$ is the characteristic orbital period at each particle's energy. The weights of each orbit are then optimized to satisfy the density and kinematic constraints, using an identical method to \citet{dillamore2025c}. See Section~3.3 in that paper for full details. The result of this process is a series of equilibrium Schwarzschild models with density $\rho_*$ and a range of anisotropies $\beta$.

To generate initial conditions for our simulations, we sample $10^6$ particle phase-space positions from each Schwarzschild model. We then integrate these orbits in the combined potential of the Milky Way and LMC, from $t=t_0$ to the present day at $t=0$. At $t=t_0$ the LMC is more than 350~kpc from the Milky Way (see Fig.~\ref{fig:lmc_orbit}), so its contribution to the potential is negligible. The Schwarzschild models are therefore very close to equilibrium at the beginning of the simulations. We note that the stellar halo was unlikely to be in perfect equilibrium before the LMC's infall (e.g., due to steady mass accretion or less massive satellites), but these effects are beyond the scope of this work.
\begin{figure}
  \centering
  \includegraphics[width=\columnwidth]{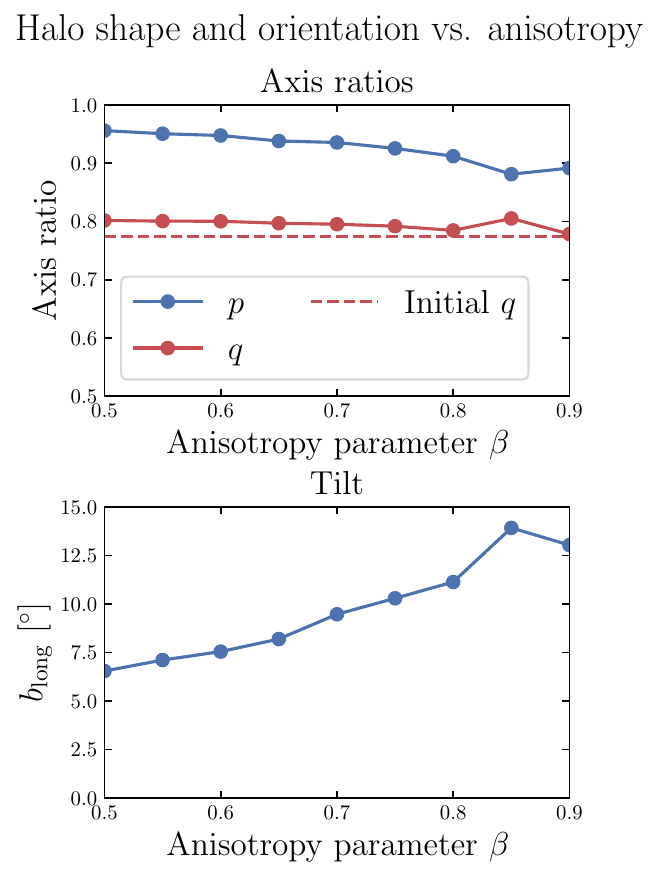}
  \caption{Final shape of the stellar haloes in our simulation as a function of anisotropy $\beta$. The fiducial simulation shown in Figs. \ref{fig:density_3d} and \ref{fig:axis_changes} is on the far right. The top and bottom panels show respectively the axis ratios and tilt of the long axis out of the Galactic plane. The dashed line in the top panel indicates the initial short-to-long axis ratio $q$ of each simulation (the initial value of $p$ is 1). As $\beta$ increases from 0.5 to 0.9, the final intermediate-to-long axis ratio decreases (the halo becomes less axisymmetric), and the tilt increases from $\approx6^\circ$ to $\approx13^\circ$.}
   \label{fig:shape_beta}
\end{figure}
\begin{figure*}
  \centering
  \includegraphics[width=\textwidth]{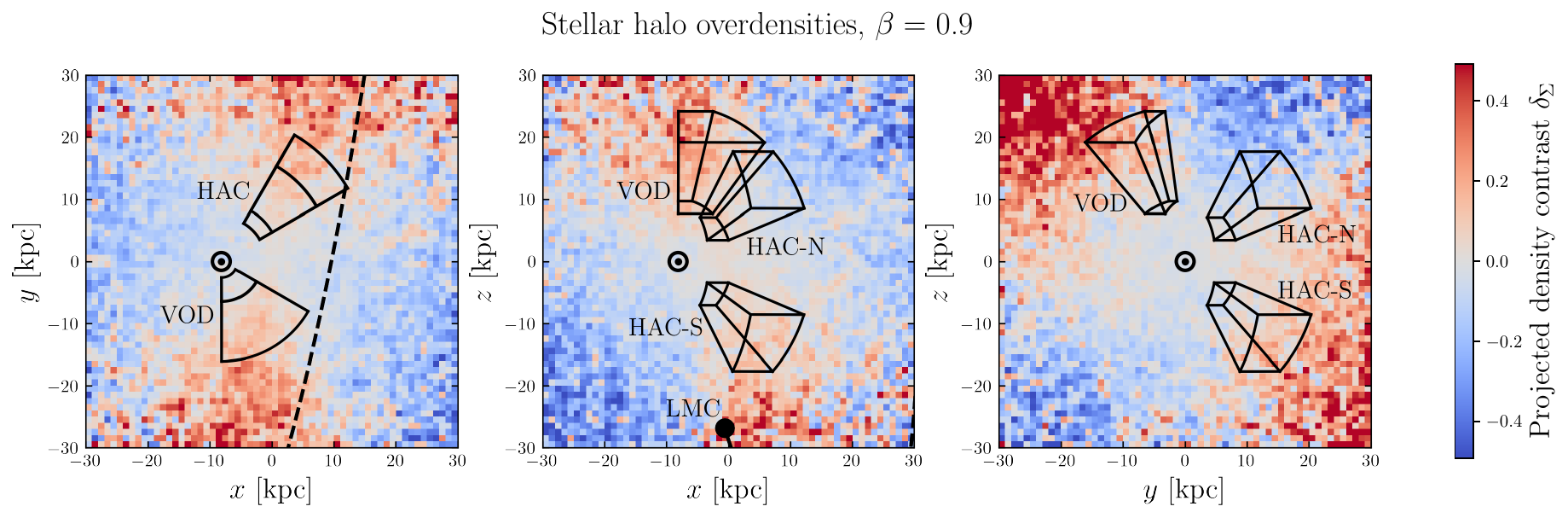}
  \caption{Fractional excess density compared to an azimuthally averaged and $z$-reflection symmetric density distribution for the $\beta=0.9$ simulation. The three panels show three different projections. The black dashed line marks the orbit of the LMC, and the location of the Sun is marked with the $\odot$ symbol. The 3D HAC and VOD regions (defined in Table \ref{tab:overdensities}) are shown with wireframes.}
   \label{fig:overdensity_3d}
\end{figure*}
\begin{figure*}
  \centering
  \includegraphics[width=0.95\textwidth]{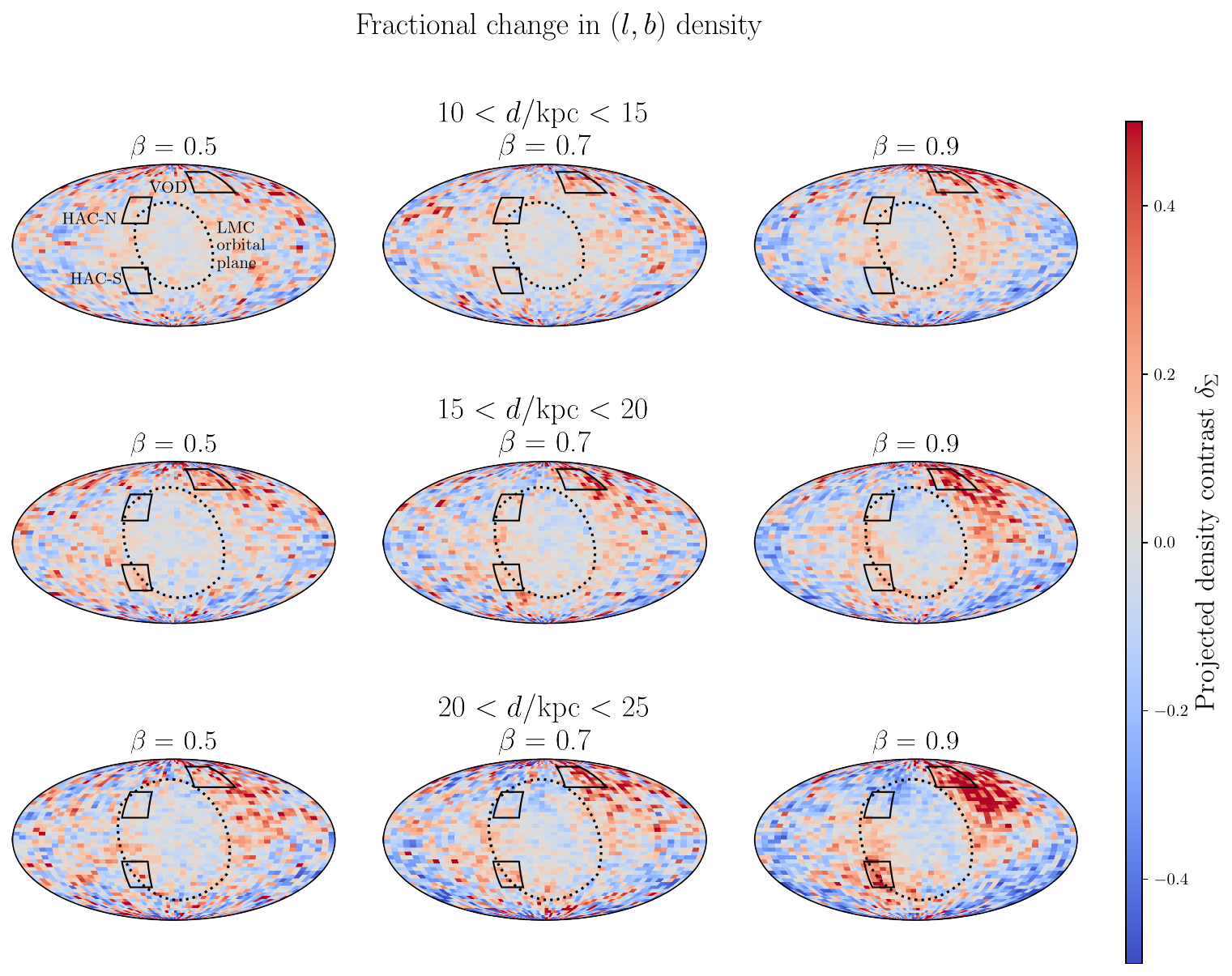}
  \caption{Fractional difference in the on-sky density between the simulated stellar halo and its azimuthal average. Each row shows a different range of heliocentric distances, and each column shows a different velocity anisotropy $\beta$. The projections are in Galactic coordinates, with $(l,b)=(0,0)$ in the centre and $l$ increasing to the left. The black boxes indicate the Hercules-Aquila Cloud (HAC) and Virgo Overdensity (VOD), as labelled in the top-left panel. The black dotted lines illustrate the orbital plane of the LMC, projected onto the median distance $d$ in each row. Strong overdensities emerge in the HAC and VOD regions of the sky when $\beta=0.9$.}
   \label{fig:density_sky_beta}
\end{figure*}
\begin{figure}
  \centering
  \includegraphics[width=\columnwidth]{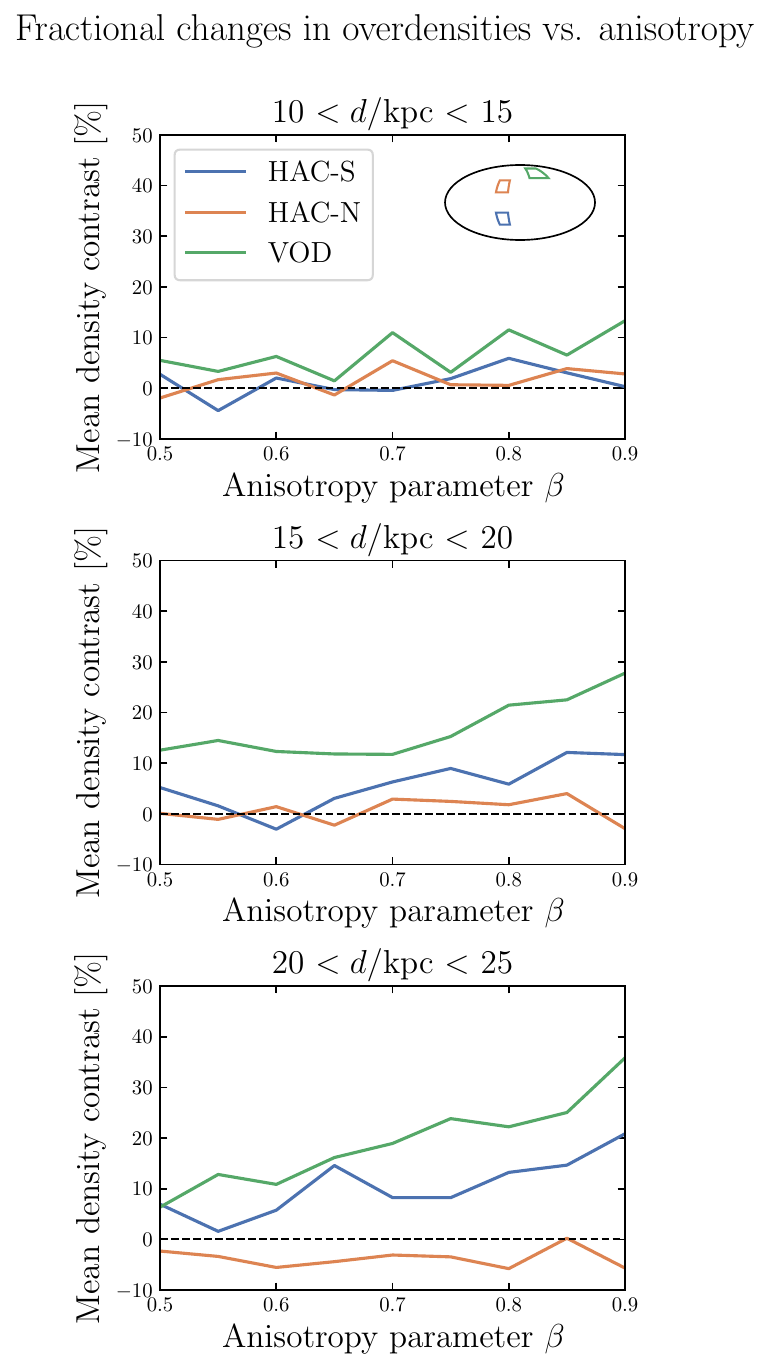}
  \caption{Fractional excess (compared to the azimuthal average) of the mean density of stars in the 3D HAC and VOD regions as a function of stellar halo anisotropy. Each panel shows a different distance range. For the HAC-S and VOD the overdensity increases as $\beta$ increases, especially at $d>15$\,kpc.}
   \label{fig:overdensity_beta}
\end{figure}
\begin{figure}
  \centering
  \includegraphics[width=\columnwidth]{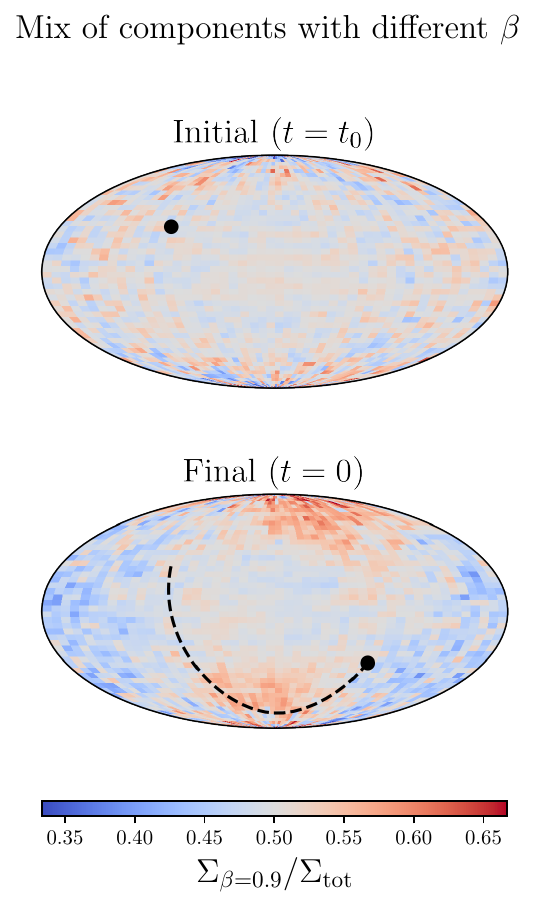}
  \caption{Fraction of on-sky density (in the distance range $10<d/\mathrm{kpc}<25$) contributed by a halo component with $\beta=0.9$. The initial distribution is made up of two components with different anisotropies $\beta=\{0.5,0.9\}$, but the same initial density distribution. The top and bottom panels show the initial (at $t=t_0$) and final (present-day) distributions. While the two populations initially have equal contributions across the sky, the $\beta=0.9$ component becomes concentrated in two overdensities after the passage of the LMC.}
   \label{fig:unmix_sky}
\end{figure}
\begin{figure*}
  \centering
  \includegraphics[width=\textwidth]{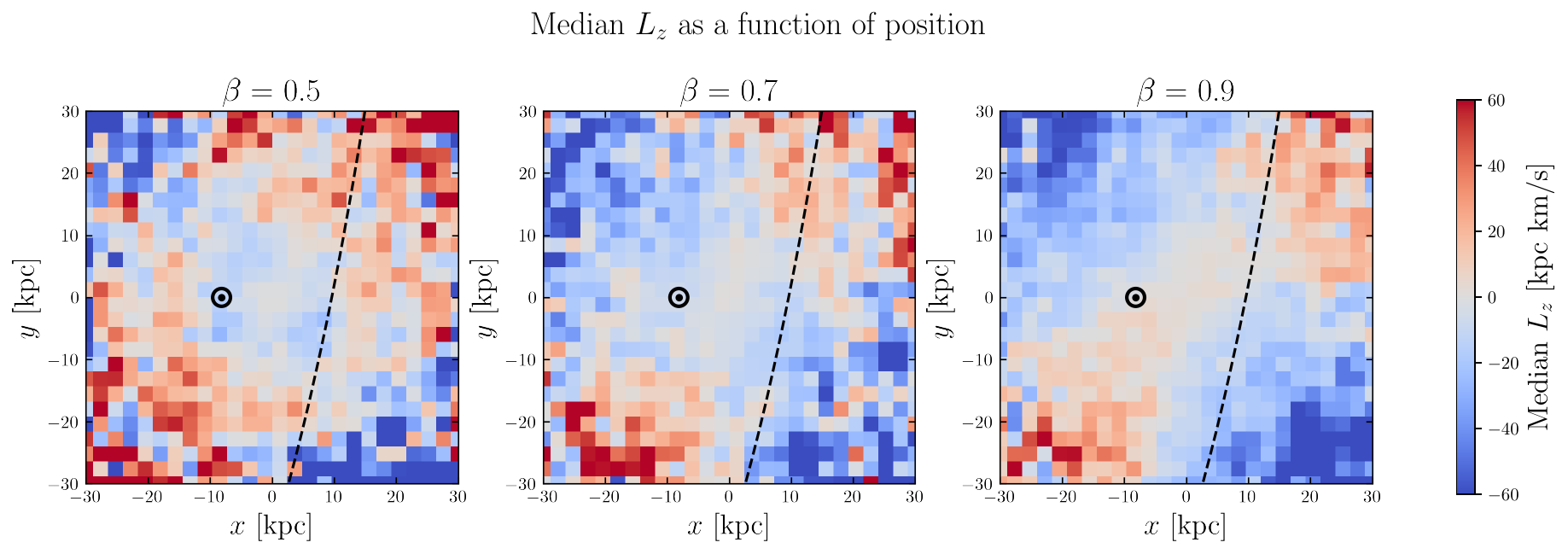}
  \caption{Median angular momentum as a function of position in the Galaxy. Each panel shows a simulation with different anisotropy $\beta$. The median is taken across all stars in a column $-30<z/\mathrm{kpc}<30$. The dashed line and $\odot$ indicate the LMC's orbit and the Sun respectively. As $\beta$ increases, a quadrupole pattern of positive and negative median $L_z$ emerges, roughly aligned with the LMC's orbital plane.}
   \label{fig:median_Lz}
\end{figure*}
\section{Results}\label{section:results}
In this section we present the results of our simulations. We split the analysis into three parts, addressing separately the global shape (Section \ref{section:shape}), individual overdensities (Section \ref{section:overdensities}), and kinematics (Section \ref{section:kinematics}).
\subsection{Shape and orientation}\label{section:shape}
We first investigate how the overall shape and orientation of the stellar halo respond to the infall of the LMC. In Fig.~\ref{fig:density_3d} we show the projected density of the fiducial $\beta=0.9$ stellar halo in Galactocentric coordinates (as defined in Fig. \ref{fig:coords}) at the beginning and end of the simulation. The Sun, the LMC, and its orbit are marked for comparison. The left-hand column (top-down view of the Galactic plane) shows that the initially axisymmetric halo becomes elongated along the direction of the LMC's orbit. This is consistent with our expectations from Section~\ref{section:theory} that highly eccentric orbits align with the LMC's tidal field in projection onto the Galactic plane. The second and third columns show edge-on projections of the Galactic plane. These show that the halo becomes tilted with respect to the Milky Way's disc. This is most obvious in the $y$-$z$ projection (bottom-right hand panel), which is close to the LMC's orbital plane. The long axis is tipped towards the past orbit of the LMC, but away from its current location. The global tilting and reshaping is distinct from the localised density wake of the LMC \citep{garavito-camargo2019}, which is visible around its orbit in the $x$-$z$ projection (bottom-middle panel). These effects also occur on different timescales; the wake is transient, while the reshaping is due to changes in individual orbits over $\sim1$\,Gyr (see Section~\ref{section:theory}).

We now quantify how the halo's global shape and orientation are affected by the LMC. As in \citet{dillamore2025c}, we estimate the shape from the inertia tensor
\begin{align}\label{eq:inertia}
    I_{ij}=\sum_{ij}^{6<r/\mathrm{kpc}<60}x_i x_j.
\end{align}
Here $x_i$ are the Galactocentric Cartesian components of position $\boldsymbol{x}$, and the sum is over all particles between radii of $r=6$ and $60$\,kpc \citep[this is the range fitted by][]{han2022}. The axis orientations are given by the eigenvectors of $I_{ij}$, and the axis ratios are equal to the ratios of the square roots of the eigenvalues. We refer to the intermediate-to-long and short-to-long axis ratios defined in this way as $p$ and $q$ respectively; note that the latter is not exactly equal to $q_*$ used to define the halo shape in Equation \eqref{eq:r_*}.

We show how these quantities evolve in our fiducial simulation (with $\beta=0.9$) in Fig.~\ref{fig:axis_changes}. The top rows show the direction of the long axis (i.e. the eigenvector with the largest eigenvalue) as a function of time, with its Galactic longitude and latitude (as viewed from the Galactic centre) on the left and right respectively. We choose to show the half of the axis in the longitude range $180^\circ\leq l<360^\circ$. The longitude is initially undefined (because the halo is initially axisymmetric), but it rapidly settles around $l_\mathrm{long}\approx260^\circ$ when perturbed by the LMC. This is approximately the orientation of the LMC's orbital plane. The latitude remains close to zero for most of the simulation, but increases to $b_\mathrm{long}\approx13^\circ$ when the LMC falls within $\sim200$ kpc of the Milky Way in the last $\sim1$\,Gyr. The LMC therefore causes the stellar halo to tilt out of the Galactic plane.

The on-sky track of the long axis is shown in Galactic coordinates in the bottom-left panel. The path traced out by the Milky Way-LMC orbit is shown with a dashed line to illustrate the LMC's orbital plane, and the final directions of the long axis and LMC are shown with circles. This shows clearly that the long axis aligns with the LMC's orbital plane. The crosses and error bars show the directions of the long axes in observationally derived stellar halo density distributions. These come from \citet{han2022} and \citet{lane2023}, both of whom fit their models to GSE stars only, and \citet{li2025}, who fit a general stellar halo sample. The final direction of the long axis in our simulation is in very close agreement with that of the \citet{lane2023} fit (though we note that their tilt $b_\mathrm{long}$ is also consistent with zero). This suggests that the LMC may be able to fully explain the tilt of the GSE debris out of the Galactic plane, removing the need for a long-lived tilted dark matter halo \citep{han2022b,dillamore2025c}. However, the long axes of the \citet{han2022} and \citet{li2025} haloes are at different longitudes, away from the LMC's orbital plane, and greater latitudes. These halo orientations therefore cannot be reproduced by our simulations with the LMC and an initially axisymmetric halo. The differences between these studies may be due to different survey footprints, distance ranges probed, or sample selections.

The bottom-right panel of Fig.~\ref{fig:axis_changes} shows the axis ratios as functions of time compared to the values and uncertainties of the \citet{han2022} fit. While the short-to-long axis ratio $q$ remains roughly constant throughout the simulation, the intermediate-to-long ratio $p$ decreases from 1 to $\sim0.9$. The LMC therefore reshapes the halo from axisymmetric to triaxial (and misaligned with the Galactic disc). However, the distribution remains closer to axisymmetric than the \citet{han2022} fit, with the final value of $p$ outside their $1\sigma$ uncertainty. The same is true for the \citet{lane2023} fit (which has $p=0.55$). Hence the LMC is not able to reshape an axisymmetric halo enough to match observations of the GSE debris. This hints that the initial shape may have needed to be triaxial to explain the current configuration. In any case, these results show that perturbations from the LMC should be corrected for \citep[e.g. see][]{correa_magnus2022} when constructing equilibrium models of the stellar halo, at least when the geometry of the potential is being constrained. This includes the Schwarzschild models of the tilted GSE debris introduced by \citet{dillamore2025c}.

We show how the final shape and tilt vary with halo anisotropy in Fig.~\ref{fig:shape_beta}. Each simulation has identical initial conditions except for the initial $\beta$. The top panel shows the axis ratios $p$ and $q$. While the short-to-long axis ratio $q$ has only weak dependence on anisotropy, the intermediate-to-long ratio $p$ decreases from $\approx0.95$ to $\approx0.89$ as $\beta$ increases from 0.5 to 0.9. Higher anisotropy therefore results in the halo becoming less axisymmetric. This is consistent with our theoretical predictions from Section~\ref{section:theory} that high-eccentricity orbits tend to be aligned.

We measure the tilt from the angle between the long axis and the Galactic plane (i.e. $b_\mathrm{long}$ in Fig.~\ref{fig:axis_changes}). This is shown in the lower panel of Fig.~\ref{fig:shape_beta}. The tilt increases with $\beta$, from $\approx7^\circ$ at $\beta=0.5$ to $\approx14^\circ$ at $\beta=0.85$. We note that the tilt at $\beta=0.9$ is slightly less than at $\beta=0.85$. It is not clear whether this is a general result, or just a symptom of the specific initial conditions used.

The reshaping and reorienting of the stellar halo caused by the LMC is therefore a strong function of anisotropy. The effects on a highly radially anisotropic component such as the GSE debris are thus likely to be much more significant than on a more isotropic distribution.
\begingroup
\begin{table}
    \centering
    \caption{Our definitions of the HAC and VOD regions. These are based on those of \citet{perottoni2022}.}
    \renewcommand{\arraystretch}{1.5}
    \begin{tabular}{c|c|c|c}
    \hline
        Region & Longitude & Latitude & Distance\\
        \hline
        HAC-S & $30^\circ<l<60^\circ$ & $-45^\circ<b<-20^\circ$ & $10<d/\mathrm{kpc}<25$\\
        HAC-N & $30^\circ<l<60^\circ$ & $20^\circ<b<45^\circ$ & $10<d/\mathrm{kpc}<25$\\
        VOD & $270^\circ<l<330^\circ$ & $50^\circ<b<75^\circ$ & $10<d/\mathrm{kpc}<25$\\
    \end{tabular}
    \label{tab:overdensities}
\end{table}
\endgroup
\subsection{Overdensities}\label{section:overdensities}
We now compare the changes in stellar halo density to observed regions of overdensity in the Milky Way's halo. We focus on the HAC and VOD. These regions are defined in Galactic coordinates in Table~\ref{tab:overdensities}, and shown on the sky in the top panel of Fig.~\ref{fig:lmc_orbit}. The $l$ and $b$ ranges are taken from \citet{perottoni2022}, but we extend the maximum distance to $d=25$\,kpc to include the full observed range of both overdensities \citep{sesar2010,simion2014,grillmair2016}. Following previous works we split the HAC into regions north and south of the Galactic plane (HAC-N and HAC-S), due to the difficulties of observing the halo at low latitude. We note that \citet{simion2014} find a much stronger overdensity in the HAC-S region compared to HAC-N.

To quantify how much the LMC creates observable substructure in the stellar halo, we require some measure of overdensity. We choose to define it as the fractional deviation from a more symmetric version of the density distribution. An equilibrium stellar halo in an axisymmetric potential has an axisymmetric density distribution with reflection symmetry in the $z=0$ plane. We therefore quantify overdensities as deviations from an azimuthally averaged density distribution which is also even in $z$. Mathematically these are described by the density contrast,
\begin{align}
    \delta_\rho(\boldsymbol{x})&\equiv\frac{\rho(\boldsymbol{x})-\bar{\rho}(R,z)}{\bar{\rho}(R,z)},\\
    \bar{\rho}(R,z)&\equiv\frac{1}{2\pi}\int_0^{2\pi}\frac{1}{2}\left[\rho(R,\phi,z)+\rho(R,\phi,-z)\right]\,\mathrm{d}\phi,
\end{align}
where the integrand in the second line is the even component of the density distribution in $z$. In practice we generate the distribution $\bar{\rho}$ by randomising the azimuths $\phi$ and signs $\mathrm{sgn}(z)$ of each particle in the simulated halo. We similarly denote $\delta_\Sigma$ as the contrast in the projected density, defined analogously to $\delta_\rho$.

In Fig.~\ref{fig:overdensity_3d} we show the projected density contrast of the fiducial simulation (with $\beta=0.9$) in Galactocentric coordinates. The densities are integrated across the cube of side length 60\,kpc. The Sun's location is indicated with the $\odot$ symbol, the Milky Way disc in grey, and the orbit of the LMC with a dashed line. We also show 3D regions corresponding to the VOD and the two parts of the HAC. As expected from Fig.~\ref{fig:axis_changes}, the LMC's reshaping of the stellar halo causes overdensities (relative to the symmetrised halo) to form. Following expectations from the analytical model in Section~\ref{section:theory}, these are aligned in azimuth with the LMC's orbital plane (left-hand panel). The right-hand panel shows that the overdensity at $y<0$ is above the Galactic plane ($z>0$), while the one at $y>0$ straddles the plane. In each projection the locations of the overdensities closely match the regions occupied by the VOD and HAC (particularly HAC-S, below the plane). This close correspondence suggests that the LMC may be at least partially responsible for creating the observed overdensities. However, we do not see any significant overdensity in the HAC-N region. This is likely due to the past orbit of the LMC in the simulation, and a different realistic trajectory \citep[e.g. see][]{vasiliev2023} may result in overdensities in different locations. We also note that some works \citep[e.g.][]{simion2014} report a density excess in HAC-S but not HAC-N, so there is not necessarily a tension between our simulation and observations.

We show the contrast in the on-sky projected density in Fig.~\ref{fig:density_sky_beta}. Each row shows a different range of distances, and each column shows a different simulation. These have anisotropies $\beta=\{0.5,0.7,0.9\}$. The theory in Section~\ref{section:theory} suggests that tidal alignment with the LMC's orbital plane should only occur for highly eccentric orbits. We therefore expect that the overdensities seen in Fig.~\ref{fig:overdensity_3d} will weaken if the anisotropy $\beta$ is reduced from the fiducial value of 0.9. Note that $\beta>0$ in each simulation, so all are radially biased.

As predicted, the general trend is that the strength of the overdensities increases as $\beta$ increases. At $\beta=0.9$ there are strong excesses in the VOD and HAC-S regions (the latter of which is weaker), but these are barely visible at $\beta=0.5$. The overdensities also increase in magnitude with increasing distance, but remain strong at $d<20$\,kpc for the $\beta=0.9$ simulation. This suggests that reshaping of the inner halo by the LMC has been underestimated by previous studies which have only considered values of $\beta$ up to 0.5 \citep[e.g.][]{garavito-camargo2019,conroy2021,rozier2022}. Furthermore the increase in density contrast with distance (also seen in Fig.~\ref{fig:overdensity_3d}) gives a prediction that the HAC and VOD overdensities should extend to larger distances than previously observed.

Fig.~\ref{fig:density_sky_beta} also predicts the presence of overdensities outside the VOD and HAC regions. Most notably, the overdensity coincident with the VOD extends to lower latitudes and below the disc. In the Milky Way these regions are partially occupied by the disc and the LMC itself, so they would be difficult to detect. At $(l,b)\sim(285^\circ,-60^\circ)$ there is another observed overdensity at a similar distance, known as the Eridanus-Phoenix (EriPhe) overdensity \citep{li2016}. It is possible that this is also linked to the HAC and VOD \citep{donlon2019,balbinot2021}, so it may be possible to associate EriPhe with the predicted overdensities in our simulations. We leave an exploration of this to a future study.

We show the anisotropy dependence of the overdensities more clearly in Fig.~\ref{fig:overdensity_beta}. We calculate the mean projected density contrast within the HAC-S, HAC-N, and VOD regions in the three distance ranges, for simulations with a range of anisotropies $\beta$. We plot these overdensity strengths as a function of $\beta$. Each panel shows a different distance range, and each colour shows a different on-sky region. This confirms the results from Fig.~\ref{fig:density_sky_beta}; as $\beta$ increases, the strengths of the HAC-S and VOD overdensities increase. They also increase in strength as distance increases. The VOD is the strongest of the overdensities, reaching a percentage excess of $\sim30-40\%$ at $d>15$\,kpc and $\beta=0.9$. This is comparable to the strength of the VOD reported by \citet{bonaca2012}. The large overdensity in the VOD region compared to the HAC may result from the reflex motion of the Milky Way's disc relative to the halo \citep[e.g.]{garavito-camargo2019,erkal2020,peterson2020}, an effect which is not captured by our analytic model. The HAC-S is roughly half as strong as the VOD, while HAC-N is much weaker or even underdense. It is therefore difficult to form an overdensity in the HAC north of the Galactic plane with the LMC. However, the evidence for an observed overdensity in the HAC-N region is weaker than in HAC-S; \citet{simion2014} find only a strong overdensity below the plane. This case is consistent with the HAC being generated by the LMC via our proposed mechanism.

\subsection{Kinematics}\label{section:kinematics}
We now address the kinematic signatures of the LMC's perturbations. Figures~\ref{fig:density_sky_beta} and \ref{fig:overdensity_beta} show that stellar haloes with identical initial density distributions but different anisotropies $\beta$ respond very differently. A corollary of this is that a stellar halo composed of multiple components with the same density distributions but different anisotropy will spatially separate when perturbed by the LMC. We demonstrate this by combining our simulations with $\beta=0.5$ and $\beta=0.9$. In Fig.~\ref{fig:unmix_sky} we plot the fraction of the on-sky surface density contributed by the $\beta=0.9$ component, at the beginning (top panel) and end (bottom) of the simulation. We include stars with heliocentric distances between $d=10$ and 20\,kpc. The LMC and its orbit are shown in black. The two populations initially have the same spatial distribution, with the surface density fraction $\Sigma_{\beta=0.9}/\Sigma_\mathrm{tot}=0.5$ everywhere. However, by the end of the simulation the $\beta=0.9$ population is more highly concentrated in two regions on the sky. These align with the past orbit of the LMC and its antipodes. In these regions the ratio reaches values of $\Sigma_{\beta=0.9}/\Sigma_\mathrm{tot}\approx2/3$, so the number of stars from the higher anisotropy population is about double that of the lower anisotropy component. The stellar halo has therefore been `fractionated' into its components with different kinematics. We have checked that the overall anisotropy profile does not change significantly as a result of the LMC.

Our analytical model in Section~\ref{section:theory} shows that the torque from the LMC acts in a quadrupole pattern, $\dot{L}_z\propto\mathrm{sin}\,2\phi$ (Equation~\ref{Lz_dot}). We therefore expect that the average angular momentum will show a similar quadrupole pattern as a function of azimuth. In Fig.~\ref{fig:median_Lz} we show the median $L_z$ as a function of Galactocentric coordinates $(x,y)$, for the simulations with anisotropies $\beta\in\{0.5,0.7,0.9\}$. The predicted quadrupole pattern does indeed appear; the top-right and bottom-left of each panel has median $L_z>0$, and the other quadrants have median $L_z<0$. This is consistent with the sign of $\dot{L}_z$ in the analytical model. The pattern becomes more pronounced and less noisy as $\beta$ increases. In particular, at radii $R<10$\,kpc it only appears at $\beta=0.9$. We therefore predict that this quadrupole pattern in $L_z$ should be present in the Solar neighbourhood, but only among stars on highly eccentric orbits (e.g. the GSE debris).

\section{Conclusions}\label{section:conclusions}
We have investigated the perturbations of the LMC on stellar haloes with different velocity anisotropies $\beta$. Unlike previous works, we have considered values of $\beta$ up to 0.9, which is realistic for the Gaia Sausage-Enceladus (GSE) debris. We have found that the high anisotropy results in much more significant changes to the inner stellar halo ($r<30$\,kpc) than previously thought. Our conclusions are summarised below.
\begin{enumerate}[label=\textbf{(\roman*)}]
    \item \textbf{Highly eccentric orbits ($e\gtrsim0.95$) are predicted by an analytical model to align preferentially with the orbital plane of the LMC due to tidal forces.} This includes the majority of stars originating in GSE.

    \item \textbf{The tidal alignment of highly eccentric orbits causes a high-$\beta$ stellar halo to be reshaped and aligned with the LMC's orbital plane.} We run a test-particle simulation of a GSE debris-like stellar halo with $\beta=0.9$ in a realistic Milky Way potential, subject to perturbations from the LMC. The initially axisymmetric halo becomes triaxial, with axis ratios changing from approximately $1:1:0.77$ to $1:0.89:0.77$. The direction of the long axis is raised out of the plane by $\approx13^\circ$. The LMC may therefore be partially responsible for the triaxiality and tilt of the stellar halo \citep{han2022,lane2023,li2025}. The long axis orientation closely matches that of the fit to the GSE debris by \citet{lane2023}, though the shape remains closer to axisymmetric than measured by \citet{han2022} or \citet{lane2023}. Hence a triaxial or tilted dark matter halo \citep{han2022b,dillamore2025c} may still be required to fully explain the present-day structure of the stellar halo.

    \item \textbf{As $\beta$ is reduced, the reshaping and realignment become less dramatic.} When $\beta=0.5$ the intermediate-to-long axis ratio is 0.96, so the halo remains closer to axisymmetric. The tilt of the long axis out of the plane is also reduced to $\approx6^\circ$. We suggest that the reshaping of the inner stellar halo was previously underestimated by studies which did not considered high GSE debris-like anisotropies ($\beta\approx0.9$).

    \item \textbf{The reshaping of the stellar halo creates on-sky projected overdensities of $\approx40\%$ at distances as close as $\sim20$\,kpc.} These are significantly closer than previously predicted, and are distinct from the well-known wake of the LMC \citep[e.g.][]{garavito-camargo2019,erkal2020,conroy2021}. We find that the overdensities become significantly weaker as $\beta$ is reduced.
    
    \item \textbf{The overdensities strongly overlap in space with the Virgo Overdensity \citep[VOD;][]{vivas2001} and the Hercules-Aquila Cloud \citep[HAC;][]{belokurov2007}.} We propose that these overdensities are not the result of unmixed debris, but instead are formed by the tidal realignment of highly eccentric orbits by the LMC. In this case they are only $\sim1$\,Gyr old and do not preserve information about the merger geometry. While a link between the VOD and LMC has previously been suggested by \citet{boubert2019}, our proposed scenario is novel. We note that we only reproduce a strong overdensity in the part of the HAC in the southern Galactic hemisphere, whereas some studies report an observed overdensity in both hemispheres. However, the exact locations of the tidally-induced overdensites likely depend on the past orbit of the LMC, which is still somewhat uncertain \citep{vasiliev2023}.

    \item \textbf{The LMC can cause different components of the halo to \textit{fractionate} (separate spatially).} This is due to the anisotropy dependence of the response. Haloes with identical initial density distributions but different kinematics are reshaped and reoriented differently, so their final distributions are distinct. In particular, a component with high anisotropy $\beta=0.9$ will be more concentrated on two regions of the sky related to the LMC's past orbit.

    \item \textbf{The LMC is predicted to introduce a quadrupole signature in the median angular momentum $L_z$ as a function of Galactic azimuth.} This is due to torquing of stellar orbits by tidal forces. This effect is apparent even at small radii, near the Sun. This may offer a new route to constraining the LMC's interaction with the Milky Way.
\end{enumerate}

In this work we have used only test particle simulations in which the Milky Way and LMC are represented by rigid potentials. In reality, the Milky Way's potential will deform in response to the infalling LMC, due to the orbits of dark matter particles being perturbed \citep[e.g.][]{lilleengen2023}. This may affect the strength or geometry of the HAC and VOD features seen in the simulations. Furthermore, a high-anisotropy dark matter halo is likely to respond in a similar way to the stellar halo. As a result, the tidal alignment effect studied in this paper may directly change the shape of the halo and thus the Milky Way's potential. It is possible that this could also affect the dynamical friction on the satellite, and thus the reconstruction of its past orbit. The response of high-$\beta$ dark matter haloes to satellite interactions therefore warrants further investigation.

We have demonstrated the importance of considering realistic kinematics when modelling the dynamics of the stellar halo. The reshaping and tilting of high-anisotropy populations by the LMC should be corrected for in future equilibrium models of the halo. These phenomena also offer a promising route to constraining the properties of the Milky Way-LMC interaction, since they affect nearby stars with higher-quality distance and velocity measurements. Upcoming observations by Rubin-LSST and spectroscopic surveys such as WEAVE will be instrumental in constraining this next generation of models.

\section*{Acknowledgements}
We are grateful to the anonymous referee for a helpful report. We thank Eugene Vasiliev and Denis Erkal for useful discussions during this project. AMD and JLS acknowledge support from the Royal Society (URF\textbackslash R1\textbackslash191555; URF\textbackslash R\textbackslash 241030).
\section*{Data Availability}
The code used in this project can be found at \url{https://github.com/adllmr/schwarzschild_lmc}. Simulation outputs are available on request.



\bibliographystyle{mnras}
\bibliography{refs} 

\begin{thebibliography}{}
\makeatletter
\relax
\def\mn@urlcharsother{\let\do\@makeother \do\$\do\&\do\#\do\^\do\_\do\%\do\~}
\def\mn@doi{\begingroup\mn@urlcharsother \@ifnextchar [ {\mn@doi@} {\mn@doi@[]}}
\def\mn@doi@[#1]#2{\def\@tempa{#1}\ifx\@tempa\@empty \href {http://dx.doi.org/#2} {doi:#2}\else \href {http://dx.doi.org/#2} {#1}\fi \endgroup}
\def\mn@eprint#1#2{\mn@eprint@#1:#2::\@nil}
\def\mn@eprint@arXiv#1{\href {http://arxiv.org/abs/#1} {{\tt arXiv:#1}}}
\def\mn@eprint@dblp#1{\href {http://dblp.uni-trier.de/rec/bibtex/#1.xml} {dblp:#1}}
\def\mn@eprint@#1:#2:#3:#4\@nil{\def\@tempa {#1}\def\@tempb {#2}\def\@tempc {#3}\ifx \@tempc \@empty \let \@tempc \@tempb \let \@tempb \@tempa \fi \ifx \@tempb \@empty \def\@tempb {arXiv}\fi \@ifundefined {mn@eprint@\@tempb}{\@tempb:\@tempc}{\expandafter \expandafter \csname mn@eprint@\@tempb\endcsname \expandafter{\@tempc}}}

\bibitem[\protect\citeauthoryear{{Astropy Collaboration} et~al.,}{{Astropy Collaboration} et~al.}{2013}]{astropy:2013}
{Astropy Collaboration} et~al., 2013, \mn@doi [\aap] {10.1051/0004-6361/201322068}, \href {http://adsabs.harvard.edu/abs/2013A%26A...558A..33A} {558, A33}

\bibitem[\protect\citeauthoryear{{Astropy Collaboration} et~al.,}{{Astropy Collaboration} et~al.}{2018}]{astropy:2018}
{Astropy Collaboration} et~al., 2018, \mn@doi [\aj] {10.3847/1538-3881/aabc4f}, \href {https://ui.adsabs.harvard.edu/abs/2018AJ....156..123A} {156, 123}

\bibitem[\protect\citeauthoryear{{Balbinot} \& {Helmi}}{{Balbinot} \& {Helmi}}{2021}]{balbinot2021}
{Balbinot} E.,  {Helmi} A.,  2021, \mn@doi [\aap] {10.1051/0004-6361/202141015}, \href {https://ui.adsabs.harvard.edu/abs/2021A&A...654A..15B} {654, A15}

\bibitem[\protect\citeauthoryear{{Belokurov} et~al.,}{{Belokurov} et~al.}{2007}]{belokurov2007}
{Belokurov} V.,  et~al., 2007, \mn@doi [\apjl] {10.1086/513144}, \href {https://ui.adsabs.harvard.edu/abs/2007ApJ...657L..89B} {657, L89}

\bibitem[\protect\citeauthoryear{{Belokurov}, {Erkal}, {Evans}, {Koposov}  \& {Deason}}{{Belokurov} et~al.}{2018}]{belokurov2018}
{Belokurov} V.,  {Erkal} D.,  {Evans} N.~W.,  {Koposov} S.~E.,   {Deason} A.~J.,  2018, \mn@doi [\mnras] {10.1093/mnras/sty982}, \href {https://ui.adsabs.harvard.edu/abs/2018MNRAS.478..611B} {478, 611}

\bibitem[\protect\citeauthoryear{{Belokurov}, {Sanders}, {Fattahi}, {Smith}, {Deason}, {Evans}  \& {Grand}}{{Belokurov} et~al.}{2020}]{belokurov2020}
{Belokurov} V.,  {Sanders} J.~L.,  {Fattahi} A.,  {Smith} M.~C.,  {Deason} A.~J.,  {Evans} N.~W.,   {Grand} R. J.~J.,  2020, \mn@doi [\mnras] {10.1093/mnras/staa876}, \href {https://ui.adsabs.harvard.edu/abs/2020MNRAS.494.3880B} {494, 3880}

\bibitem[\protect\citeauthoryear{{Belokurov}, {Vasiliev}, {Deason}, {Koposov}, {Fattahi}, {Dillamore}, {Davies}  \& {Grand}}{{Belokurov} et~al.}{2023}]{belokurov_chevrons}
{Belokurov} V.,  {Vasiliev} E.,  {Deason} A.~J.,  {Koposov} S.~E.,  {Fattahi} A.,  {Dillamore} A.~M.,  {Davies} E.~Y.,   {Grand} R. J.~J.,  2023, \mn@doi [\mnras] {10.1093/mnras/stac3436}, \href {https://ui.adsabs.harvard.edu/abs/2023MNRAS.518.6200B} {518, 6200}

\bibitem[\protect\citeauthoryear{{Bennett} \& {Bovy}}{{Bennett} \& {Bovy}}{2019}]{bennett2019}
{Bennett} M.,  {Bovy} J.,  2019, \mn@doi [\mnras] {10.1093/mnras/sty2813}, \href {https://ui.adsabs.harvard.edu/abs/2019MNRAS.482.1417B} {482, 1417}

\bibitem[\protect\citeauthoryear{{Binney} \& {Tremaine}}{{Binney} \& {Tremaine}}{2008}]{binney_tremaine}
{Binney} J.,  {Tremaine} S.,  2008, {Galactic Dynamics: Second Edition}

\bibitem[\protect\citeauthoryear{{Bird}, {Xue}, {Liu}, {Shen}, {Flynn}, {Yang}, {Zhao}  \& {Tian}}{{Bird} et~al.}{2021}]{bird2021}
{Bird} S.~A.,  {Xue} X.-X.,  {Liu} C.,  {Shen} J.,  {Flynn} C.,  {Yang} C.,  {Zhao} G.,   {Tian} H.-J.,  2021, \mn@doi [\apj] {10.3847/1538-4357/abfa9e}, \href {https://ui.adsabs.harvard.edu/abs/2021ApJ...919...66B} {919, 66}

\bibitem[\protect\citeauthoryear{{Bonaca} et~al.,}{{Bonaca} et~al.}{2012}]{bonaca2012}
{Bonaca} A.,  et~al., 2012, \mn@doi [\aj] {10.1088/0004-6256/143/5/105}, \href {https://ui.adsabs.harvard.edu/abs/2012AJ....143..105B} {143, 105}

\bibitem[\protect\citeauthoryear{{Bonaca} et~al.,}{{Bonaca} et~al.}{2020}]{bonaca2020}
{Bonaca} A.,  et~al., 2020, \mn@doi [\apjl] {10.3847/2041-8213/ab9caa}, \href {https://ui.adsabs.harvard.edu/abs/2020ApJ...897L..18B} {897, L18}

\bibitem[\protect\citeauthoryear{{Boubert}, {Belokurov}, {Erkal}  \& {Iorio}}{{Boubert} et~al.}{2019}]{boubert2019}
{Boubert} D.,  {Belokurov} V.,  {Erkal} D.,   {Iorio} G.,  2019, \mn@doi [\mnras] {10.1093/mnras/sty3014}, \href {https://ui.adsabs.harvard.edu/abs/2019MNRAS.482.4562B} {482, 4562}

\bibitem[\protect\citeauthoryear{{Brooks}, {Sanders}, {Chandra}, {Garavito-Camargo}, {Dillamore}, {Price-Whelan}  \& {Ting}}{{Brooks} et~al.}{2025}]{brooks2025b}
{Brooks} R. A.~N.,  {Sanders} J.~L.,  {Chandra} V.,  {Garavito-Camargo} N.,  {Dillamore} A.~M.,  {Price-Whelan} A.~M.,   {Ting} Y.-S.,  2025, \mn@doi [arXiv e-prints] {10.48550/arXiv.2510.04735}, \href {https://ui.adsabs.harvard.edu/abs/2025arXiv251004735B} {p. arXiv:2510.04735}

\bibitem[\protect\citeauthoryear{{Brooks}, {Sanders}, {Dillamore}, {Garavito-Camargo}  \& {Price-Whelan}}{{Brooks} et~al.}{2026}]{Brooks2026}
{Brooks} R. A.~N.,  {Sanders} J.~L.,  {Dillamore} A.~M.,  {Garavito-Camargo} N.,   {Price-Whelan} A.~M.,  2026, \mn@doi [\mnras] {10.1093/mnras/staf2111}, \href {https://ui.adsabs.harvard.edu/abs/2026MNRAS.545f2111B} {545, staf2111}

\bibitem[\protect\citeauthoryear{{Bystr{\"o}m} et~al.,}{{Bystr{\"o}m} et~al.}{2025}]{bystrom2025}
{Bystr{\"o}m} A.,  et~al., 2025, \mn@doi [\mnras] {10.1093/mnras/staf1219}, \href {https://ui.adsabs.harvard.edu/abs/2025MNRAS.542..560B} {542, 560}

\bibitem[\protect\citeauthoryear{{Chandra} et~al.,}{{Chandra} et~al.}{2025}]{chandra2025}
{Chandra} V.,  et~al., 2025, \mn@doi [\apj] {10.3847/1538-4357/addab6}, \href {https://ui.adsabs.harvard.edu/abs/2025ApJ...988..156C} {988, 156}

\bibitem[\protect\citeauthoryear{{Chandrasekhar}}{{Chandrasekhar}}{1943}]{chandrasekhar1943}
{Chandrasekhar} S.,  1943, \mn@doi [\apj] {10.1086/144517}, \href {https://ui.adsabs.harvard.edu/abs/1943ApJ....97..255C} {97, 255}

\bibitem[\protect\citeauthoryear{{Conroy}, {Naidu}, {Garavito-Camargo}, {Besla}, {Zaritsky}, {Bonaca}  \& {Johnson}}{{Conroy} et~al.}{2021}]{conroy2021}
{Conroy} C.,  {Naidu} R.~P.,  {Garavito-Camargo} N.,  {Besla} G.,  {Zaritsky} D.,  {Bonaca} A.,   {Johnson} B.~D.,  2021, \mn@doi [\nat] {10.1038/s41586-021-03385-7}, \href {https://ui.adsabs.harvard.edu/abs/2021Natur.592..534C} {592, 534}

\bibitem[\protect\citeauthoryear{{Correa Magnus} \& {Vasiliev}}{{Correa Magnus} \& {Vasiliev}}{2022}]{correa_magnus2022}
{Correa Magnus} L.,  {Vasiliev} E.,  2022, \mn@doi [\mnras] {10.1093/mnras/stab3726}, \href {https://ui.adsabs.harvard.edu/abs/2022MNRAS.511.2610C} {511, 2610}

\bibitem[\protect\citeauthoryear{{Das}, {Hawkins}  \& {Jofr{\'e}}}{{Das} et~al.}{2020}]{das2020}
{Das} P.,  {Hawkins} K.,   {Jofr{\'e}} P.,  2020, \mn@doi [\mnras] {10.1093/mnras/stz3537}, \href {https://ui.adsabs.harvard.edu/abs/2020MNRAS.493.5195D} {493, 5195}

\bibitem[\protect\citeauthoryear{{Deason} \& {Belokurov}}{{Deason} \& {Belokurov}}{2024}]{deason2024}
{Deason} A.~J.,  {Belokurov} V.,  2024, \mn@doi [\nar] {10.1016/j.newar.2024.101706}, \href {https://ui.adsabs.harvard.edu/abs/2024NewAR..9901706D} {99, 101706}

\bibitem[\protect\citeauthoryear{{Dillamore} \& {Sanders}}{{Dillamore} \& {Sanders}}{2026}]{dillamore2025c}
{Dillamore} A.~M.,  {Sanders} J.~L.,  2026, \mn@doi [\mnras] {10.1093/mnras/stag226}, \href {https://ui.adsabs.harvard.edu/abs/2026MNRAS.546ag226D} {546, stag226}

\bibitem[\protect\citeauthoryear{{Dillamore}, {Belokurov}  \& {Evans}}{{Dillamore} et~al.}{2024}]{dillamore2024}
{Dillamore} A.~M.,  {Belokurov} V.,   {Evans} N.~W.,  2024, \mn@doi [\mnras] {10.1093/mnras/stae1789}, \href {https://ui.adsabs.harvard.edu/abs/2024MNRAS.532.4389D} {532, 4389}

\bibitem[\protect\citeauthoryear{{Donlon}, {Newberg}, {Weiss}, {Amy}  \& {Thompson}}{{Donlon} et~al.}{2019}]{donlon2019}
{Donlon} II T.,  {Newberg} H.~J.,  {Weiss} J.,  {Amy} P.,   {Thompson} J.,  2019, \mn@doi [\apj] {10.3847/1538-4357/ab4f72}, \href {https://ui.adsabs.harvard.edu/abs/2019ApJ...886...76D} {886, 76}

\bibitem[\protect\citeauthoryear{{Donlon}, {Newberg}, {Sanderson}, {Bregou}, {Horta}, {Arora}  \& {Panithanpaisal}}{{Donlon} et~al.}{2024}]{donlon2023}
{Donlon} T.,  {Newberg} H.~J.,  {Sanderson} R.,  {Bregou} E.,  {Horta} D.,  {Arora} A.,   {Panithanpaisal} N.,  2024, \mn@doi [\mnras] {10.1093/mnras/stae1264}, \href {https://ui.adsabs.harvard.edu/abs/2024MNRAS.531.1422D} {531, 1422}

\bibitem[\protect\citeauthoryear{{Erkal} et~al.,}{{Erkal} et~al.}{2019}]{erkal2019}
{Erkal} D.,  et~al., 2019, \mn@doi [\mnras] {10.1093/mnras/stz1371}, \href {https://ui.adsabs.harvard.edu/abs/2019MNRAS.487.2685E} {487, 2685}

\bibitem[\protect\citeauthoryear{{Erkal}, {Belokurov}  \& {Parkin}}{{Erkal} et~al.}{2020}]{erkal2020}
{Erkal} D.,  {Belokurov} V.~A.,   {Parkin} D.~L.,  2020, \mn@doi [\mnras] {10.1093/mnras/staa2840}, \href {https://ui.adsabs.harvard.edu/abs/2020MNRAS.498.5574E} {498, 5574}

\bibitem[\protect\citeauthoryear{{Erkal} et~al.,}{{Erkal} et~al.}{2021}]{erkal2021}
{Erkal} D.,  et~al., 2021, \mn@doi [\mnras] {10.1093/mnras/stab1828}, \href {https://ui.adsabs.harvard.edu/abs/2021MNRAS.506.2677E} {506, 2677}

\bibitem[\protect\citeauthoryear{{Feuillet}, {Feltzing}, {Sahlholdt}  \& {Casagrande}}{{Feuillet} et~al.}{2020}]{feuillet2020}
{Feuillet} D.~K.,  {Feltzing} S.,  {Sahlholdt} C.~L.,   {Casagrande} L.,  2020, \mn@doi [\mnras] {10.1093/mnras/staa1888}, \href {https://ui.adsabs.harvard.edu/abs/2020MNRAS.497..109F} {497, 109}

\bibitem[\protect\citeauthoryear{{Feuillet}, {Sahlholdt}, {Feltzing}  \& {Casagrande}}{{Feuillet} et~al.}{2021}]{feuillet2021}
{Feuillet} D.~K.,  {Sahlholdt} C.~L.,  {Feltzing} S.,   {Casagrande} L.,  2021, \mn@doi [\mnras] {10.1093/mnras/stab2614}, \href {https://ui.adsabs.harvard.edu/abs/2021MNRAS.508.1489F} {508, 1489}

\bibitem[\protect\citeauthoryear{{Foote} et~al.,}{{Foote} et~al.}{2023}]{foote2023}
{Foote} H.~R.,  et~al., 2023, \mn@doi [\apj] {10.3847/1538-4357/ace533}, \href {https://ui.adsabs.harvard.edu/abs/2023ApJ...954..163F} {954, 163}

\bibitem[\protect\citeauthoryear{{GRAVITY Collaboration} et~al.,}{{GRAVITY Collaboration} et~al.}{2018}]{gravity2018}
{GRAVITY Collaboration} et~al., 2018, \mn@doi [\aap] {10.1051/0004-6361/201833718}, \href {https://ui.adsabs.harvard.edu/abs/2018A&A...615L..15G} {615, L15}

\bibitem[\protect\citeauthoryear{{Gaia Collaboration} et~al.,}{{Gaia Collaboration} et~al.}{2016}]{gaia}
{Gaia Collaboration} et~al., 2016, \mn@doi [\aap] {10.1051/0004-6361/201629272}, \href {https://ui.adsabs.harvard.edu/abs/2016A&A...595A...1G} {595, A1}

\bibitem[\protect\citeauthoryear{{Gaia Collaboration} et~al.,}{{Gaia Collaboration} et~al.}{2021}]{luri2021}
{Gaia Collaboration} et~al., 2021, \mn@doi [\aap] {10.1051/0004-6361/202039588}, \href {https://ui.adsabs.harvard.edu/abs/2021A&A...649A...7G} {649, A7}

\bibitem[\protect\citeauthoryear{{Garavito-Camargo}, {Besla}, {Laporte}, {Johnston}, {G{\'o}mez}  \& {Watkins}}{{Garavito-Camargo} et~al.}{2019}]{garavito-camargo2019}
{Garavito-Camargo} N.,  {Besla} G.,  {Laporte} C. F.~P.,  {Johnston} K.~V.,  {G{\'o}mez} F.~A.,   {Watkins} L.~L.,  2019, \mn@doi [\apj] {10.3847/1538-4357/ab32eb}, \href {https://ui.adsabs.harvard.edu/abs/2019ApJ...884...51G} {884, 51}

\bibitem[\protect\citeauthoryear{{Garavito-Camargo}, {Besla}, {Laporte}, {Price-Whelan}, {Cunningham}, {Johnston}, {Weinberg}  \& {G{\'o}mez}}{{Garavito-Camargo} et~al.}{2021}]{garavito-camargo2021}
{Garavito-Camargo} N.,  {Besla} G.,  {Laporte} C. F.~P.,  {Price-Whelan} A.~M.,  {Cunningham} E.~C.,  {Johnston} K.~V.,  {Weinberg} M.,   {G{\'o}mez} F.~A.,  2021, \mn@doi [\apj] {10.3847/1538-4357/ac0b44}, \href {https://ui.adsabs.harvard.edu/abs/2021ApJ...919..109G} {919, 109}

\bibitem[\protect\citeauthoryear{{Gibbons}, {Belokurov}  \& {Evans}}{{Gibbons} et~al.}{2017}]{gibbons2017}
{Gibbons} S.~L.~J.,  {Belokurov} V.,   {Evans} N.~W.,  2017, \mn@doi [\mnras] {10.1093/mnras/stw2328}, \href {https://ui.adsabs.harvard.edu/abs/2017MNRAS.464..794G} {464, 794}

\bibitem[\protect\citeauthoryear{{Grillmair} \& {Carlin}}{{Grillmair} \& {Carlin}}{2016}]{grillmair2016}
{Grillmair} C.~J.,  {Carlin} J.~L.,  2016, in {Newberg} H.~J.,  {Carlin} J.~L.,  eds,  Astrophysics and Space Science Library Vol. 420, Tidal Streams in the Local Group and Beyond. p.~87 (\mn@eprint {arXiv} {1603.08936}), \mn@doi{10.1007/978-3-319-19336-6_4}

\bibitem[\protect\citeauthoryear{{Han} et~al.,}{{Han} et~al.}{2022a}]{han2022}
{Han} J.~J.,  et~al., 2022a, \mn@doi [\aj] {10.3847/1538-3881/ac97e9}, \href {https://ui.adsabs.harvard.edu/abs/2022AJ....164..249H} {164, 249}

\bibitem[\protect\citeauthoryear{{Han} et~al.,}{{Han} et~al.}{2022b}]{han2022b}
{Han} J.~J.,  et~al., 2022b, \mn@doi [\apj] {10.3847/1538-4357/ac795f}, \href {https://ui.adsabs.harvard.edu/abs/2022ApJ...934...14H} {934, 14}

\bibitem[\protect\citeauthoryear{{Han}, {Semenov}, {Conroy}  \& {Hernquist}}{{Han} et~al.}{2023}]{han2023}
{Han} J.~J.,  {Semenov} V.,  {Conroy} C.,   {Hernquist} L.,  2023, \mn@doi [\apjl] {10.3847/2041-8213/ad0641}, \href {https://ui.adsabs.harvard.edu/abs/2023ApJ...957L..24H} {957, L24}

\bibitem[\protect\citeauthoryear{{Helmi}, {Babusiaux}, {Koppelman}, {Massari}, {Veljanoski}  \& {Brown}}{{Helmi} et~al.}{2018}]{helmi2018}
{Helmi} A.,  {Babusiaux} C.,  {Koppelman} H.~H.,  {Massari} D.,  {Veljanoski} J.,   {Brown} A. G.~A.,  2018, \mn@doi [\nat] {10.1038/s41586-018-0625-x}, \href {https://ui.adsabs.harvard.edu/abs/2018Natur.563...85H} {563, 85}

\bibitem[\protect\citeauthoryear{{Henon}}{{Henon}}{1959}]{Henon}
{Henon} M.,  1959, Annales d'Astrophysique, \href {https://ui.adsabs.harvard.edu/abs/1959AnAp...22..126H} {22, 126}

\bibitem[\protect\citeauthoryear{{Iorio} \& {Belokurov}}{{Iorio} \& {Belokurov}}{2019}]{iorio2019}
{Iorio} G.,  {Belokurov} V.,  2019, \mn@doi [\mnras] {10.1093/mnras/sty2806}, \href {https://ui.adsabs.harvard.edu/abs/2019MNRAS.482.3868I} {482, 3868}

\bibitem[\protect\citeauthoryear{{Iorio} \& {Belokurov}}{{Iorio} \& {Belokurov}}{2021}]{iorio2021}
{Iorio} G.,  {Belokurov} V.,  2021, \mn@doi [\mnras] {10.1093/mnras/stab005}, \href {https://ui.adsabs.harvard.edu/abs/2021MNRAS.502.5686I} {502, 5686}

\bibitem[\protect\citeauthoryear{{Koposov} et~al.,}{{Koposov} et~al.}{2023}]{koposov2023}
{Koposov} S.~E.,  et~al., 2023, \mn@doi [\mnras] {10.1093/mnras/stad551}, \href {https://ui.adsabs.harvard.edu/abs/2023MNRAS.521.4936K} {521, 4936}

\bibitem[\protect\citeauthoryear{{Lancaster}, {Koposov}, {Belokurov}, {Evans}  \& {Deason}}{{Lancaster} et~al.}{2019}]{lancaster2019}
{Lancaster} L.,  {Koposov} S.~E.,  {Belokurov} V.,  {Evans} N.~W.,   {Deason} A.~J.,  2019, \mn@doi [\mnras] {10.1093/mnras/stz853}, \href {https://ui.adsabs.harvard.edu/abs/2019MNRAS.486..378L} {486, 378}

\bibitem[\protect\citeauthoryear{{Lane}, {Bovy}  \& {Mackereth}}{{Lane} et~al.}{2023}]{lane2023}
{Lane} J. M.~M.,  {Bovy} J.,   {Mackereth} J.~T.,  2023, \mn@doi [\mnras] {10.1093/mnras/stad2834}, \href {https://ui.adsabs.harvard.edu/abs/2023MNRAS.526.1209L} {526, 1209}

\bibitem[\protect\citeauthoryear{{Laporte} \& {Orkney}}{{Laporte} \& {Orkney}}{2026}]{laporte2026}
{Laporte} C. F.~P.,  {Orkney} M. D.~A.,  2026, \mn@doi [arXiv e-prints] {10.48550/arXiv.2604.14502}, \href {https://ui.adsabs.harvard.edu/abs/2026arXiv260414502L} {p. arXiv:2604.14502}

\bibitem[\protect\citeauthoryear{{Law} \& {Majewski}}{{Law} \& {Majewski}}{2010}]{law2010}
{Law} D.~R.,  {Majewski} S.~R.,  2010, \mn@doi [\apj] {10.1088/0004-637X/714/1/229}, \href {https://ui.adsabs.harvard.edu/abs/2010ApJ...714..229L} {714, 229}

\bibitem[\protect\citeauthoryear{{Li} et~al.,}{{Li} et~al.}{2016}]{li2016}
{Li} T.~S.,  et~al., 2016, \mn@doi [\apj] {10.3847/0004-637X/817/2/135}, \href {https://ui.adsabs.harvard.edu/abs/2016ApJ...817..135L} {817, 135}

\bibitem[\protect\citeauthoryear{{Li} et~al.,}{{Li} et~al.}{2026}]{li2025}
{Li} S.,  et~al., 2026, \mn@doi [\apj] {10.3847/1538-4357/ae41b9}, \href {https://ui.adsabs.harvard.edu/abs/2026ApJ...999..108L} {999, 108}

\bibitem[\protect\citeauthoryear{{Lilleengen} et~al.,}{{Lilleengen} et~al.}{2023}]{lilleengen2023}
{Lilleengen} S.,  et~al., 2023, \mn@doi [\mnras] {10.1093/mnras/stac3108}, \href {https://ui.adsabs.harvard.edu/abs/2023MNRAS.518..774L} {518, 774}

\bibitem[\protect\citeauthoryear{{Lucey}, {Mateu}, {Price-Whelan}, {Hogg}, {Rix}  \& {Sanderson}}{{Lucey} et~al.}{2026}]{lucey2025}
{Lucey} M.,  {Mateu} C.,  {Price-Whelan} A.~M.,  {Hogg} D.~W.,  {Rix} H.-W.,   {Sanderson} R.~E.,  2026, \mn@doi [\aj] {10.3847/1538-3881/ae4aaa}, \href {https://ui.adsabs.harvard.edu/abs/2026AJ....171..249L} {171, 249}

\bibitem[\protect\citeauthoryear{{Lynden-Bell}}{{Lynden-Bell}}{1963}]{lynden-bell1963}
{Lynden-Bell} D.,  1963, The Observatory, \href {https://ui.adsabs.harvard.edu/abs/1963Obs....83...23L} {83, 23}

\bibitem[\protect\citeauthoryear{{Mackereth} \& {Bovy}}{{Mackereth} \& {Bovy}}{2020}]{mackereth2020}
{Mackereth} J.~T.,  {Bovy} J.,  2020, \mn@doi [\mnras] {10.1093/mnras/staa047}, \href {https://ui.adsabs.harvard.edu/abs/2020MNRAS.492.3631M} {492, 3631}

\bibitem[\protect\citeauthoryear{{McMillan}}{{McMillan}}{2017}]{mcmillan17}
{McMillan} P.~J.,  2017, \mn@doi [\mnras] {10.1093/mnras/stw2759}, \href {https://ui.adsabs.harvard.edu/abs/2017MNRAS.465...76M} {465, 76}

\bibitem[\protect\citeauthoryear{{Naidu}, {Conroy}, {Bonaca}, {Johnson}, {Ting}, {Caldwell}, {Zaritsky}  \& {Cargile}}{{Naidu} et~al.}{2020}]{naidu2020}
{Naidu} R.~P.,  {Conroy} C.,  {Bonaca} A.,  {Johnson} B.~D.,  {Ting} Y.-S.,  {Caldwell} N.,  {Zaritsky} D.,   {Cargile} P.~A.,  2020, \mn@doi [\apj] {10.3847/1538-4357/abaef4}, \href {https://ui.adsabs.harvard.edu/abs/2020ApJ...901...48N} {901, 48}

\bibitem[\protect\citeauthoryear{{Naidu} et~al.,}{{Naidu} et~al.}{2021}]{naidu2021}
{Naidu} R.~P.,  et~al., 2021, \mn@doi [\apj] {10.3847/1538-4357/ac2d2d}, \href {https://ui.adsabs.harvard.edu/abs/2021ApJ...923...92N} {923, 92}

\bibitem[\protect\citeauthoryear{{Navarro}, {Frenk}  \& {White}}{{Navarro} et~al.}{1997}]{NFW}
{Navarro} J.~F.,  {Frenk} C.~S.,   {White} S. D.~M.,  1997, \mn@doi [\apj] {10.1086/304888}, \href {https://ui.adsabs.harvard.edu/abs/1997ApJ...490..493N} {490, 493}

\bibitem[\protect\citeauthoryear{{Newberg} et~al.,}{{Newberg} et~al.}{2002}]{newberg2002}
{Newberg} H.~J.,  et~al., 2002, \mn@doi [\apj] {10.1086/338983}, \href {https://ui.adsabs.harvard.edu/abs/2002ApJ...569..245N} {569, 245}

\bibitem[\protect\citeauthoryear{{Pe{\~n}arrubia}, {G{\'o}mez}, {Besla}, {Erkal}  \& {Ma}}{{Pe{\~n}arrubia} et~al.}{2016}]{penarrubia2016}
{Pe{\~n}arrubia} J.,  {G{\'o}mez} F.~A.,  {Besla} G.,  {Erkal} D.,   {Ma} Y.-Z.,  2016, \mn@doi [\mnras] {10.1093/mnrasl/slv160}, \href {https://ui.adsabs.harvard.edu/abs/2016MNRAS.456L..54P} {456, L54}

\bibitem[\protect\citeauthoryear{{Perottoni}, {Limberg}, {Amarante}, {Rossi}, {Queiroz}, {Santucci}, {P{\'e}rez-Villegas}  \& {Chiappini}}{{Perottoni} et~al.}{2022}]{perottoni2022}
{Perottoni} H.~D.,  {Limberg} G.,  {Amarante} J. A.~S.,  {Rossi} S.,  {Queiroz} A. B.~A.,  {Santucci} R.~M.,  {P{\'e}rez-Villegas} A.,   {Chiappini} C.,  2022, \mn@doi [\apjl] {10.3847/2041-8213/ac88d6}, \href {https://ui.adsabs.harvard.edu/abs/2022ApJ...936L...2P} {936, L2}

\bibitem[\protect\citeauthoryear{{Petersen} \& {Pe{\~n}arrubia}}{{Petersen} \& {Pe{\~n}arrubia}}{2020}]{peterson2020}
{Petersen} M.~S.,  {Pe{\~n}arrubia} J.,  2020, \mn@doi [\mnras] {10.1093/mnrasl/slaa029}, \href {https://ui.adsabs.harvard.edu/abs/2020MNRAS.494L..11P} {494, L11}

\bibitem[\protect\citeauthoryear{{Pietrzy{\'n}ski} et~al.,}{{Pietrzy{\'n}ski} et~al.}{2019}]{pietrzynski2019}
{Pietrzy{\'n}ski} G.,  et~al., 2019, \mn@doi [\nat] {10.1038/s41586-019-0999-4}, \href {https://ui.adsabs.harvard.edu/abs/2019Natur.567..200P} {567, 200}

\bibitem[\protect\citeauthoryear{{Read} \& {Erkal}}{{Read} \& {Erkal}}{2019}]{read2019}
{Read} J.~I.,  {Erkal} D.,  2019, \mn@doi [\mnras] {10.1093/mnras/stz1320}, \href {https://ui.adsabs.harvard.edu/abs/2019MNRAS.487.5799R} {487, 5799}

\bibitem[\protect\citeauthoryear{{Rozier}, {Famaey}, {Siebert}, {Monari}, {Pichon}  \& {Ibata}}{{Rozier} et~al.}{2022}]{rozier2022}
{Rozier} S.,  {Famaey} B.,  {Siebert} A.,  {Monari} G.,  {Pichon} C.,   {Ibata} R.,  2022, \mn@doi [\apj] {10.3847/1538-4357/ac7139}, \href {https://ui.adsabs.harvard.edu/abs/2022ApJ...933..113R} {933, 113}

\bibitem[\protect\citeauthoryear{{Sch{\"o}nrich}, {Binney}  \& {Dehnen}}{{Sch{\"o}nrich} et~al.}{2010}]{schonrich2010}
{Sch{\"o}nrich} R.,  {Binney} J.,   {Dehnen} W.,  2010, \mn@doi [\mnras] {10.1111/j.1365-2966.2010.16253.x}, \href {https://ui.adsabs.harvard.edu/abs/2010MNRAS.403.1829S} {403, 1829}

\bibitem[\protect\citeauthoryear{{Schwarzschild}}{{Schwarzschild}}{1979}]{schwarzschild1979}
{Schwarzschild} M.,  1979, \mn@doi [\apj] {10.1086/157282}, \href {https://ui.adsabs.harvard.edu/abs/1979ApJ...232..236S} {232, 236}

\bibitem[\protect\citeauthoryear{{Sesar} et~al.,}{{Sesar} et~al.}{2010}]{sesar2010}
{Sesar} B.,  et~al., 2010, \mn@doi [\apj] {10.1088/0004-637X/708/1/717}, \href {https://ui.adsabs.harvard.edu/abs/2010ApJ...708..717S} {708, 717}

\bibitem[\protect\citeauthoryear{{Shipp} et~al.,}{{Shipp} et~al.}{2021}]{shipp2021}
{Shipp} N.,  et~al., 2021, \mn@doi [\apj] {10.3847/1538-4357/ac2e93}, \href {https://ui.adsabs.harvard.edu/abs/2021ApJ...923..149S} {923, 149}

\bibitem[\protect\citeauthoryear{{Simion}, {Belokurov}, {Irwin}  \& {Koposov}}{{Simion} et~al.}{2014}]{simion2014}
{Simion} I.~T.,  {Belokurov} V.,  {Irwin} M.,   {Koposov} S.~E.,  2014, \mn@doi [\mnras] {10.1093/mnras/stu133}, \href {https://ui.adsabs.harvard.edu/abs/2014MNRAS.440..161S} {440, 161}

\bibitem[\protect\citeauthoryear{{Simion}, {Belokurov}  \& {Koposov}}{{Simion} et~al.}{2019}]{simion2019}
{Simion} I.~T.,  {Belokurov} V.,   {Koposov} S.~E.,  2019, \mn@doi [\mnras] {10.1093/mnras/sty2744}, \href {https://ui.adsabs.harvard.edu/abs/2019MNRAS.482..921S} {482, 921}

\bibitem[\protect\citeauthoryear{{Vasiliev}}{{Vasiliev}}{2019}]{agama}
{Vasiliev} E.,  2019, \mn@doi [\mnras] {10.1093/mnras/sty2672}, \href {https://ui.adsabs.harvard.edu/abs/2019MNRAS.482.1525V} {482, 1525}

\bibitem[\protect\citeauthoryear{{Vasiliev}}{{Vasiliev}}{2023}]{vasiliev2023}
{Vasiliev} E.,  2023, \mn@doi [Galaxies] {10.3390/galaxies11020059}, \href {https://ui.adsabs.harvard.edu/abs/2023Galax..11...59V} {11, 59}

\bibitem[\protect\citeauthoryear{{Vasiliev}}{{Vasiliev}}{2024}]{vasiliev2024}
{Vasiliev} E.,  2024, \mn@doi [\mnras] {10.1093/mnras/stad2612}, \href {https://ui.adsabs.harvard.edu/abs/2024MNRAS.527..437V} {527, 437}

\bibitem[\protect\citeauthoryear{{Vasiliev} \& {Belokurov}}{{Vasiliev} \& {Belokurov}}{2020}]{vasiliev2020_sgr}
{Vasiliev} E.,  {Belokurov} V.,  2020, \mn@doi [\mnras] {10.1093/mnras/staa2114}, \href {https://ui.adsabs.harvard.edu/abs/2020MNRAS.497.4162V} {497, 4162}

\bibitem[\protect\citeauthoryear{{Vasiliev}, {Belokurov}  \& {Erkal}}{{Vasiliev} et~al.}{2021}]{vasiliev_tango}
{Vasiliev} E.,  {Belokurov} V.,   {Erkal} D.,  2021, \mn@doi [\mnras] {10.1093/mnras/staa3673}, \href {https://ui.adsabs.harvard.edu/abs/2021MNRAS.501.2279V} {501, 2279}

\bibitem[\protect\citeauthoryear{{Vivas} et~al.,}{{Vivas} et~al.}{2001}]{vivas2001}
{Vivas} A.~K.,  et~al., 2001, \mn@doi [\apjl] {10.1086/320915}, \href {https://ui.adsabs.harvard.edu/abs/2001ApJ...554L..33V} {554, L33}

\bibitem[\protect\citeauthoryear{{Weinberg}}{{Weinberg}}{1989}]{weinberg1989}
{Weinberg} M.~D.,  1989, \mn@doi [\mnras] {10.1093/mnras/239.2.549}, \href {https://ui.adsabs.harvard.edu/abs/1989MNRAS.239..549W} {239, 549}

\bibitem[\protect\citeauthoryear{{Yaaqib}, {Petersen}  \& {Pe{\~n}arrubia}}{{Yaaqib} et~al.}{2024}]{yaaqib2024}
{Yaaqib} R.,  {Petersen} M.~S.,   {Pe{\~n}arrubia} J.,  2024, \mn@doi [\mnras] {10.1093/mnras/stae1363}, \href {https://ui.adsabs.harvard.edu/abs/2024MNRAS.531.3524Y} {531, 3524}

\bibitem[\protect\citeauthoryear{{Yan}, {Shi}, {Chen}, {Zhao}  \& {Zhao}}{{Yan} et~al.}{2023}]{yan2023}
{Yan} H.~H.,  {Shi} W.~B.,  {Chen} Y.~Q.,  {Zhao} J.~K.,   {Zhao} G.,  2023, \mn@doi [\aap] {10.1051/0004-6361/202346249}, \href {https://ui.adsabs.harvard.edu/abs/2023A&A...674A..78Y} {674, A78}

\bibitem[\protect\citeauthoryear{{Ye}, {Du}, {Deng}, {Liao}, {Huang}, {Shi}  \& {Ma}}{{Ye} et~al.}{2024}]{ye2024}
{Ye} D.,  {Du} C.,  {Deng} M.,  {Liao} J.,  {Huang} Y.,  {Shi} J.,   {Ma} J.,  2024, \mn@doi [\mnras] {10.1093/mnras/stae1655}, \href {https://ui.adsabs.harvard.edu/abs/2024MNRAS.532.2584Y} {532, 2584}

\bibitem[\protect\citeauthoryear{{van der Marel} \& {Kallivayalil}}{{van der Marel} \& {Kallivayalil}}{2014}]{vandermarel2014}
{van der Marel} R.~P.,  {Kallivayalil} N.,  2014, \mn@doi [\apj] {10.1088/0004-637X/781/2/121}, \href {https://ui.adsabs.harvard.edu/abs/2014ApJ...781..121V} {781, 121}

\bibitem[\protect\citeauthoryear{{van der Marel}, {Alves}, {Hardy}  \& {Suntzeff}}{{van der Marel} et~al.}{2002}]{vandermarel2002}
{van der Marel} R.~P.,  {Alves} D.~R.,  {Hardy} E.,   {Suntzeff} N.~B.,  2002, \mn@doi [\aj] {10.1086/343775}, \href {https://ui.adsabs.harvard.edu/abs/2002AJ....124.2639V} {124, 2639}

\makeatother
\end{thebibliography}




\appendix

\section{Derivation of $\langle R^2 \rangle$ in the isochrone potential}\label{section:mean_R_squared}
In this section we derive the mean squared radius $\langle R^2 \rangle$ of an orbit in the isochrone potential as a function of its energy and angular momentum. In this potential, orbits are best described analytically in terms of a parameter $\eta$, defined by \citet{binney_tremaine} as
\begin{equation}
    s=2+\frac{c}{b}(1-e\,\mathrm{cos}\,\eta),
\end{equation}
where
\begin{align}
    c&\equiv a-b\\
    e^2&\equiv1-\frac{aL^2}{GMc^2}\label{eq:e}\\
    s&\equiv1+\sqrt{1+\frac{r^2}{b^2}},\label{eq:s}
\end{align}
and $a$ is given by Equation~\eqref{eq:semi-major_axis}. The radial angle (which increases linearly in time) is
\begin{equation}
    \theta_r=\eta-\frac{ec}{a}\mathrm{sin}\,\eta.
\end{equation}
For orbits in the plane $z=0$ (for which $r=R$ and $L^2=L_z^2$), the mean squared radius is therefore (from Equation \ref{eq:s})
\begin{align}
    \langle R^2 \rangle&=b^2\left[\left\langle(s-1)^2\right\rangle-1\right]\\
    &=b^2\left[\left\langle\left(1+\frac{c}{b}(1-e\,\mathrm{cos}\,\eta)\right)^2\right\rangle-1\right]\\
    &=2bc\left(1-e\langle\mathrm{cos}\,\eta\rangle\right)+c^2\left\langle\left(1-e\,\mathrm{cos}\,\eta\right)^2\right\rangle\\
    &=c(2b+c)-2ace\langle\mathrm{cos}\,\eta\rangle+c^2e^2\langle\mathrm{cos}^2\eta\rangle.
\end{align}
The averaged cosines are
\begin{align}
    \langle\mathrm{cos}\,\eta\rangle&=\frac{1}{\pi}\int_0^\pi\mathrm{cos}\,\eta\,\mathrm{d}\theta_r\\
    &=\frac{1}{\pi}\left[\int_0^\pi\mathrm{cos}\,\eta\,\mathrm{d}\eta-\frac{ec}{a}\int_0^\pi\mathrm{cos}^2\eta\,\mathrm{d}\eta\right]\\
    &=-\frac{ec}{2a},\\
    \langle\mathrm{cos}^2\eta\rangle&=\frac{1}{\pi}\left[\int_0^\pi\mathrm{cos}^2\eta\,\mathrm{d}\eta-\frac{ec}{a}\int_0^\pi\mathrm{cos}^3\eta\,\mathrm{d}\eta\right]\\
    &=\frac{1}{2},
\end{align}
giving the mean squared radius in terms of $L_z$ and $a$,
\begin{align}
    \langle R^2 \rangle&=c(2b+c)+\frac{3}{2}c^2e^2\\
    &=\frac{1}{2}\left[(5a-b)(a-b)-\frac{3aL_z^2}{GM}\right].\label{eq:mean_R_squared}
\end{align}
For a circular orbit, $e=0$ and hence $L_z^2=GMc^2/a$ from Equation \eqref{eq:e}. Combining with Equation \eqref{eq:mean_R_squared},
this gives the circular orbit radius in terms of $a$,
\begin{equation}
    r_\mathrm{circ}^2=a^2-b^2.
\end{equation}


\bsp	
\label{lastpage}
\end{document}